\documentclass[%
reprint,
 amsmath,amssymb,
 aps,
]{revtex4-2}

\usepackage{graphicx}
\usepackage{dcolumn}
\usepackage{bm}
\usepackage[utf8]{inputenc}
\usepackage{etoolbox}
\usepackage{siunitx}
\usepackage{float}
\usepackage{subfiles}
\usepackage{caption}
\usepackage{subcaption}
\usepackage{multirow}
\usepackage{hyperref}
\usepackage{xcolor}
\hypersetup{
    colorlinks=true,
    linkcolor=blue,
    urlcolor=blue,
    citecolor=blue,
    }
\begin{document}

\title[The P$^3$ Experiment]{The P$^3$ Experiment: \\A Positron Source Demonstrator for Future Lepton Colliders}


\author{N.~Vallis}
\altaffiliation[Also at ]{EPFL, Lausanne, Switzerland}
\author{P.~Craievich}
\author{M.~Sch\"{a}r}
\author{R.~Zennaro}
\author{B.~Auchmann}
\altaffiliation[Also at ]{CERN, Geneva, Switzerland}
\author{H.H.~Braun}
\author{M.I.~Besana}
\author{M.~Duda}
\author{R.~Fortunati}
\author{H.~Garcia-Rodrigues}
\author{D.~Hauenstein}
\author{R.~Ischebeck}
\author{E.~Ismaili}
\author{P.~Jurani\'{c}}
\author{J.~Kosse}
\author{A.~Magazinik}
\altaffiliation[Also at ]{CERN, Geneva, Switzerland}
\author{F.~Marcellini}
\author{T.U.~Michlmayr}
\author{S.~M\"{u}ller}
\author{M.~Pedrozzi}
\author{R.~Rotundo}
\author{G.L.~Orlandi}
\author{M.~Seidel}
\altaffiliation[Also at ]{EPFL, Lausanne, Switzerland}
\author{N.~Strohmaier}
\author{M.~Zykova}
\affiliation{Paul Scherrer Institut, Villigen, Switzerland}
\email{nicolas.vallis@psi.ch}
\date{\today}

\begin{abstract}
The PSI Positron Production (P$^3$ or P-cubed) experiment is a demonstrator for an electron-driven positron source and capture system with potential to improve by an order of magnitude the state-of-the-art positron yield normalized to the drive linac energy. The experiment is framed in the FCC-ee injector study and will be hosted in the SwissFEL facility at the Paul Scherrer Institute in Switzerland. This paper is an overview of the P$^3$ design at an advanced stage, with a particular emphasis on a novel positron capture system and its associated beam dynamics. Additionally, a concept for the experiment diagnostics is presented, as well as the key points of the ongoing installation works. 
\end{abstract}

\maketitle

\section{INTRODUCTION}
\label{ch:Intro}

Positron (e+) sources for particle accelerators are almost universally based on pair production through high energy electron (e-) beams impinging upon high-Z converter targets~\cite{chehab:CAS}. Despite the large e+ yields provided, such particle showers have an extreme transverse emittance and energy spread associated, making e+ beams require a significantly greater damping than an equivalent e- beam~\cite{clendenin:positrons}. The standard e+ collection systems use high-field solenoids around the production target and along a great part of the capture linac to maximize transmission up to a damping ring (DR), where e+ have their emittance cooled. However, the poor e+ capture efficiency provided by conventional solenoid focusing is the fundamental reason of low yields ---defined as the ratio between accepted e+ (N$_{e+}$) by the DR and primary e- at target (N$_{e-}$)--- present in all ever existing e+ accelerators~\cite{Chaikovska:PositronSources}.

This is particularly true for high intensity machines such as SLC~\cite{SLCHandbook} (SLAC, USA), where the all time high e+ yield was recorded, and SuperKEKB~\cite{Akai:SuperKEKB} (KEK, Japan), hosting the current state of the art e+ source. Table~\ref{tab:e+StateOfTheArt} overviews e+ production and injection at such facilities, including key parameters for e+ injection efficiency: the primary e- energy, the magnetic strength around the target and the linac, and the iris aperture of the RF capture cavities. Notice that despite the great e+ yields at the target from multi-GeV e- beams and the multi-Tesla solenoid focusing, the achieved yield at the DR is arguably low compared to the large e+ production at the target exit. 

\begin{table}[h]
\caption{e+ injection performance overview of SuperKEKB and SLC according to~\cite{Chaikovska:PositronSources}.}
\label{tab:e+StateOfTheArt}
\begin{ruledtabular}
\begin{tabular}{lcc}
		& SLC  & SuperKEKB\\
        Operation period& 1989 - 1998  &  2014 - \\
        \colrule
        Primary e- energy [GeV]                   & 30-33 & 3.5  \\
		Max. sol. field at Target [T]                  & 5.5 & 3.5  \\
		Avg. sol. field along linac [T]                & 0.5 & 0.4  \\
        Min. RF cavity aperture [mm]           &18~ & 30  \\
        e+ yield in target region\footnote{Approximate values derived from \cite{SLCHandbook} and \cite{Akai:SuperKEKB}}                        & $\approx$~30~ & $\approx$~8~\\
        Max. meas. e+ yield at DR                            & 2.5~\cite{clendenin:positrons} & 0.63~\cite{Suwada:Detection} \\
\end{tabular}
\end{ruledtabular}
\end{table}

\begin{figure*}
    \centering
    \includegraphics[width=\textwidth]{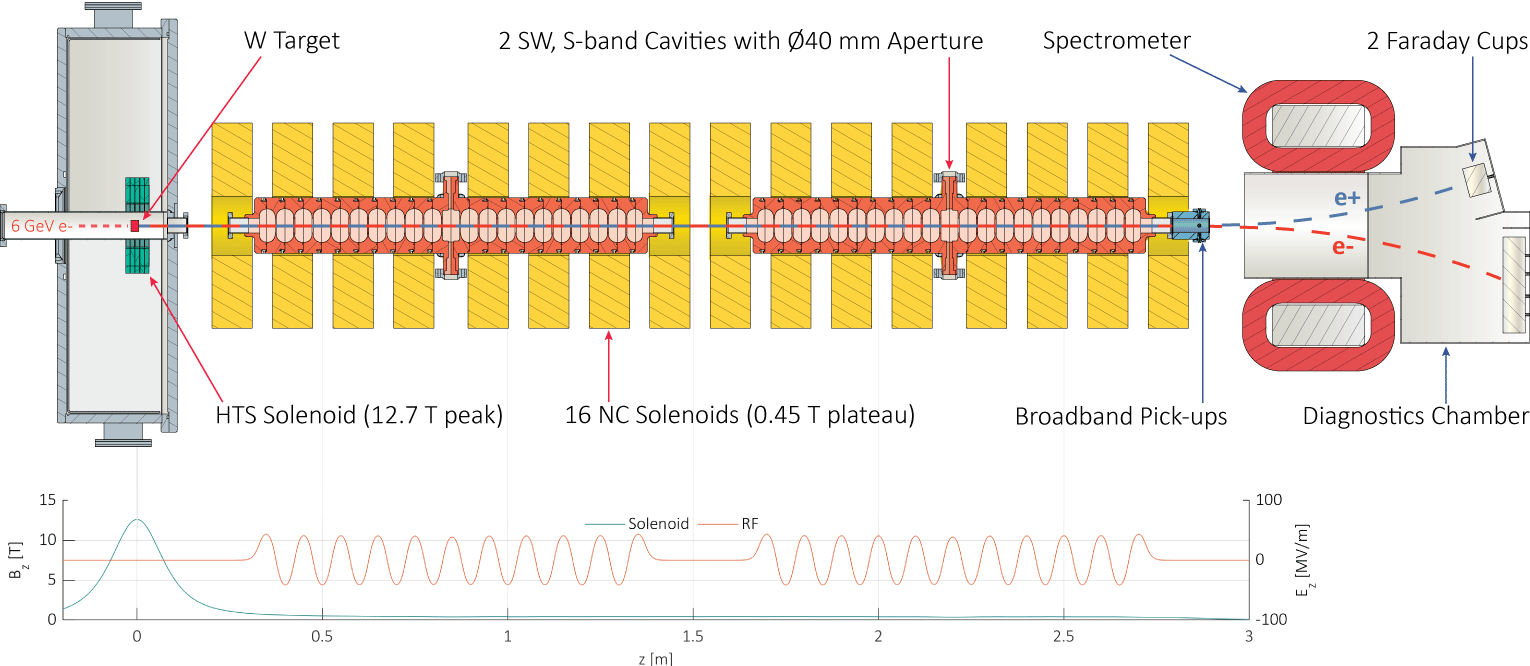}
    \caption{\label{fig:P3Layout} Simplified layout of the P$^3$ experiment featuring key components of the e+ source and capture system (red arrows) and diagnostics (blue arrows). Featuring real dimensions and solenoid and RF field plots at corresponding z.}
\end{figure*}

The SwissFEL facility~\cite{Prat:SiwssFEL} (PSI, Switzerland) will host the PSI Positron Production (P$^3$ or P-cubed) experiment, a e+ source demonstrator with potential to improve significantly the state-of-the-art e+ yield --- by an order of magnitude with respect to e+ sources driven by e- beams in a similar energy range ---. The experiment layout is shown in Fig.~\ref{fig:P3Layout}, featuring a e+ source based on a 6~GeV electron (e-) beam and 17.5~mm-thick (or 5 times the radiation length) amorphous Tungsten target, followed by a capture system consisting of a solenoid system and 2 RF accelerating cavities. The remarkable e+ capture capabilities of P$^3$ are enabled to great extent by the usage of  high temperature superconducting (HTS) solenoid around the target region, as well as a novel standing wave solution for the RF cavities that provides a large iris aperture to maximize e+ capture (see Table~\ref{tab:FCCvsP3}). In addition, a variety of beam diagnostics, whose goal is to demonstrate such a e+ yield upgrade, will detect simultaneously the captured e+e- time structure, and measure the bunch charge and energy spectrum of e+ and e- streams separately. 

The P$^3$ project is driven by the FCC-ee~\cite{FCC-ee_CDR, FCCweb} luminosity requirements and its design and results will constitute one of the main deliverables of the FCC-ee injector feasiblity study~\cite{Craievich:FCCeeInjector, CHART21, CHART22}. Although the experiment is designed to reproduce the beam dynamics of the FCC-ee e+ source~\cite{Chaikovska:FCC-ee}, the primary e- beam current parameters are relaxed mainly to meet the SwissFEL radiation protection limits. Notice in Table.~\ref{tab:FCCvsP3} the differences in bunch charge, repetition rate and the number of bunches per pulse with respect to the FCC-ee baseline. 

\begin{table}[h]
\caption{Main e+ source parameters of FCC-ee and P$^3$.}
\label{tab:FCCvsP3}
\begin{ruledtabular}
\begin{tabular}{lcc}
		& FCC-ee \cite{CHART22}  & P$^3$\\
        \colrule
		Energy [GeV]                              & \multicolumn{2}{c}{6}\\
        Max. sol. field at Target [T]                  & tbd & 12.7  \\
		Avg. sol. field along linac [T]                & 0.5  & 0.45\\
        Min. RF cavity aperture [mm]              & 60  & 40 \\
        \colrule
		$\sigma_E$ & \multicolumn{2}{c}{0.1$\%$}\\
        $\sigma_t$ [ps] & \multicolumn{2}{c}{3.33}\\
        $\sigma_x$, $\sigma_y$ [mm] &  \multicolumn{2}{c}{0.5}\\
		$\sigma_{px}$, $\sigma_{py}$ [MeV/c] & \multicolumn{2}{c}{0.06}\\
        Target length [mm] & \multicolumn{2}{c}{17.5}\\
		$Q_{bunch}$ [nC]                          & 1.7 - 2.4      & 0.20\\
		Reptition rate [Hz]                       & 200     & 1\\
		Bunches per pulse                         & 2     & 1\\
\end{tabular}
\end{ruledtabular}
\end{table}
This paper provides a comprehensive overview the P$^3$ experiment at a highly advanced design stage, the most emphasis being on the technology that will enable our novel capture system and its associated beam dynamics. In addition, the concept design of the beam diagnostics tools is discussed, introducing in some cases novel-approach solutions particularly suitable for highly spread e+e- beams. The text concludes with a brief summary of the current installation works at SwissFEL covering the dedicated extraction line, RF power source and the bunker of the P$^3$ experiment.

\section{Key Technology}
\label{ch:KeyTech}

The P$^3$ experiment will employ novel and conventional technology for e+ capture and transport from the target to the diagnostics section. As illustrated in Fig.~\ref{fig:P3Layout}, this process will rely on three key devices: a high-temperature superconducting (HTS) solenoid around the production target, 2 RF accelerating cavities immediately downstream from the target and 16 normal conducting (NC) solenoids around the RF cavities.

\subsection{HTS Solenoid}
\label{ch:HTS}
\begin{figure*}
    \centering
    \includegraphics[width=\textwidth]{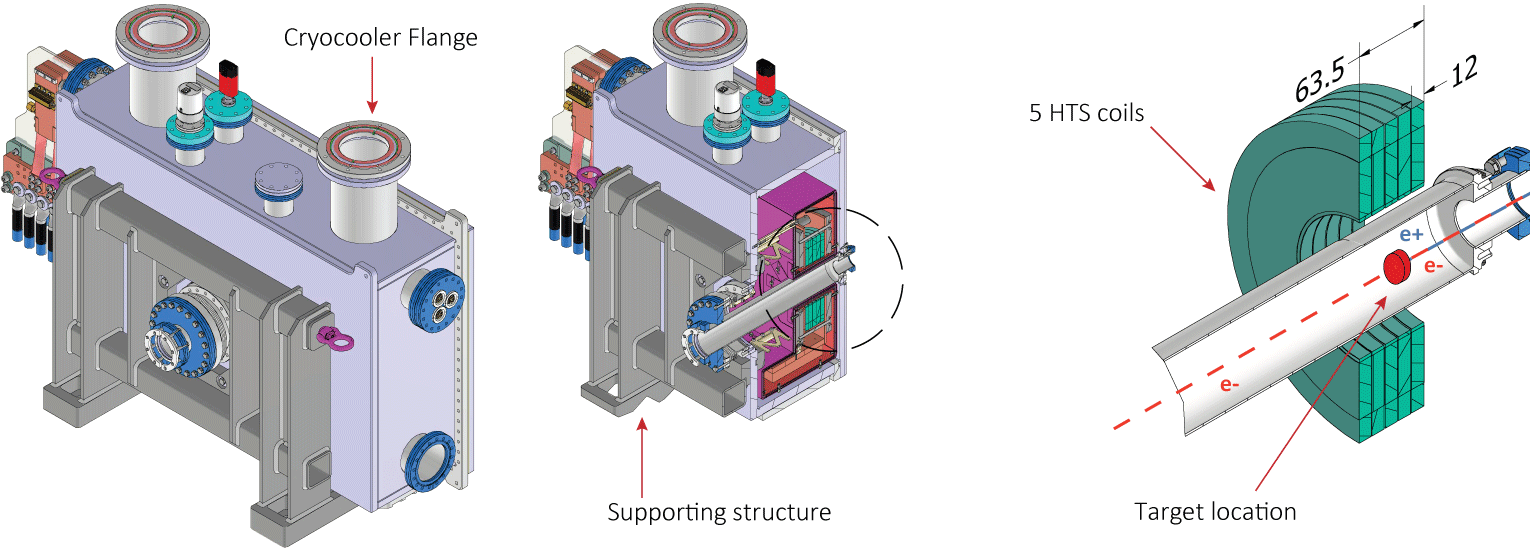}
    \caption{\label{fig:HTScoils} Assembly of the HTS solenoid, cryostat and supporting structure (left). Including section view (center) and detail view of the coils and pipe only around target area (right).}
\end{figure*}

Multi-Tesla solenoid fields around the production target are the basis of  standard e+ collection systems~\cite{chehab:CAS}. To this end, the P$^3$ experiment will use a high temperature superconducting (HTS) solenoid in order to deliver a peak 12.7~T on-axis field near the target exit face. As shown in Fig.~\ref{fig:HTScoils}, the HTS solenoid is an arrangement of 5 coils, which are made out of non-insulated (NI) ReBCO tape winded in-house at PSI. The usage of NI HTS technology enables extremely high magnetic strengths, unprecedented in other e+ sources, which increase enormously the e+ capture capabilities of conventional, normal conducting flux concentrators~\cite{Zhao:AMDcomparison, Zhao:AMDOptimization}. An analogous HTS solution is considered for FCC-ee, for which P$^3$ will serve as a demonstrator in most aspects of beam dynamics and operation. However, the lower radiation levels at SwissFEL with respect to FCC-ee (see Table \ref{tab:FCCvsP3}), allow for a great simplification of the radiation shielding design. These differences are remarkable in terms of in terms of the expected dose per year at the HTS coils (18 kGy in P$^3$ vs. 23 MGy in FCC-ee) and displaced atoms per year (1e-8 DPA vs. 2e-4 DPA) \cite{Humann:AMDRadiation}.

\begin{figure}[h!]
    \centering
    \includegraphics[width=0.45\textwidth]{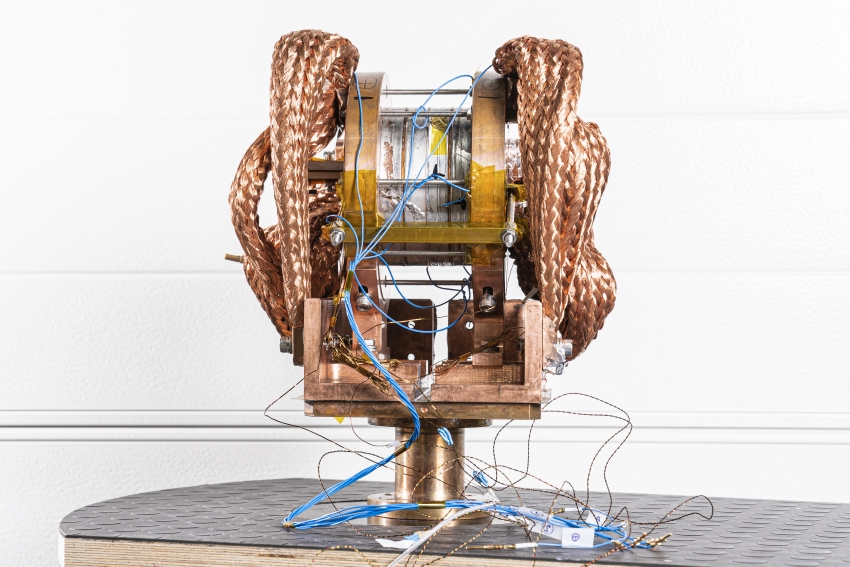}
    \caption{\label{fig:HTSPrototype} Prototype assembly of 5 HTS coils with mechanical suppot and high current leads. Photographed at PSI in June 2022.}
\end{figure}

NI HTS magnets have demonstrated great stability during high current operation \cite{Hahn:HTSNoInsulation, Hahn:HTSChallenges}. ReBCO tape allows for conduction-cooled, cryogen-free operation at 15~K where the risk of radiation-induced damage of the insulation is negligible \cite{Fischer:ReBCORadiation}. The HTS coils will sit inside a cryostat with two single-stage cryocoolers \cite{SumitomoCryoCoolers}, respectively dedicated to the coils and the radiation shield and 1.2~kA current leads. Moreover, the conventionally long charging times of NI HTS magnets are significantly reduced due to the compact size of the solenoid. A prototype of the HTS solenoid for P$^3$, shown in Fig.~\ref{fig:HTSPrototype}, has been successfully winded, soldered and stacked in-house. In addition, tests at PSI have demonstrated cryogen-free operation at 15~K and 2~kA, measuring peak magnetic fields of 18~T on-axis.  

\begin{table}[h]
\caption{Overview of HTS solenoid parameters}
\label{tab:HTSParams}
\begin{ruledtabular}
\begin{tabular}{lc}
		Conductor & ReBCO tape\\
  	Number of coils & 5\\
        Thickness [mm] & 12\\
		Coil Diameter [mm] & 122 (inner), 219 (outer)\\
		Aperture [mm] & 72\\
		Heat load [W] & 9 (at 15~K), 106 (at 40~K)\\
		Max. magnetic field [T] & 15 (on axis), 21 (in conductor)\\
		Operating current [kA] & 1.17 \\
		Charging time [h] & 11 \\
\end{tabular}
\end{ruledtabular}
\end{table}

\subsection{RF Cavities}
\label{RFCavities}

\begin{figure}[h!]
    \centering
    \includegraphics[width=0.4\textwidth]{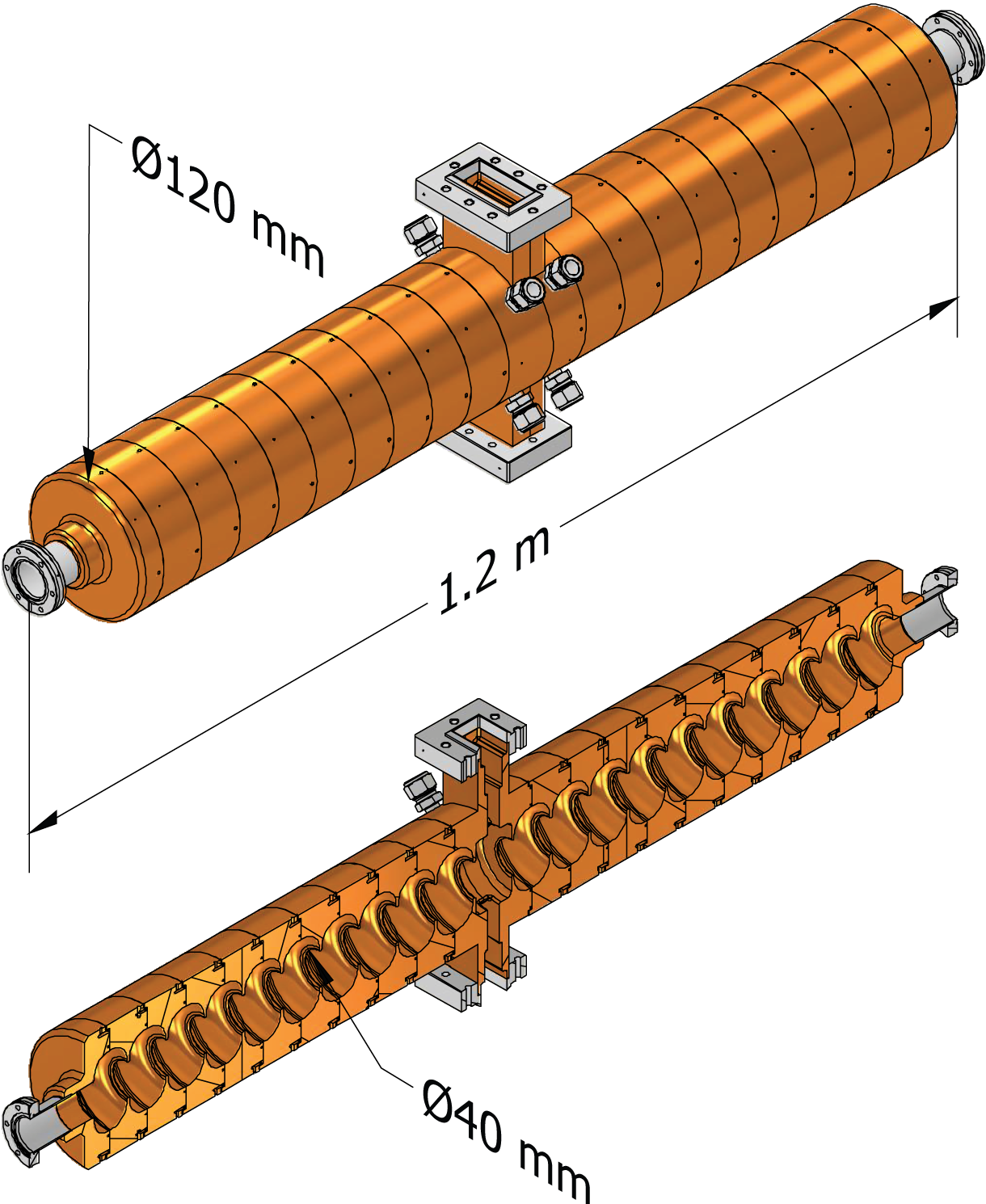}
    \caption{\label{fig:RFCavities} Mechanical design of RF cavities (top) and section view (bottom).}
\end{figure}

The e+ capture into into stable RF buckets is provided by two S-band, standing wave (SW) cavities, shown in Fig.~\ref{fig:RFCavities}, and whose parameters are listed in Table~\ref{tab:RFCavitiesParams}. A novel SW solution with a large iris aperture of 40~mm diameter will allow for an increased transverse acceptance and while maintaining a reasonably high shunt impedance. The availability of commercial klystrons and conventional waveguide components in European S-band determined the frequency choice of \SI{2.9988}{GHz}. Each cavity is connected to the waveguide network through a double feeder coupler, placed centrally in order to increase the mode separation. One single klystron modulator system, similar to those already installed at the SwissFEL linac, can provide the required peak power and RF pulse length to fill the two cavities and reach an effective grandient of 18 MV/m. In addition, the coupling factor and the total amount of cells per cavitiy have been defined to optimize the operation for a \SI{3}{\micro s} RF pulse length. While the normal repetition rates during the experiment will be low (see Table~\ref{tab:FCCvsP3}) the cavities can operate up to 100~Hz, which allows for a reduced conditioning time. The RF phases of both cavities will be adjusted independently through a high-power, in-vacuum phase shifter developed at PSI.

\begin{table}[h]
\caption{Overview of RF cavities' parameters}
\label{tab:RFCavitiesParams}
\begin{ruledtabular}
\begin{tabular}{lc}
		Length [m] & 1.2\\
		RF frequency [GHz] & 2.9988 (S-band)\\
		Nominal gradient [MV/m] & 18\\
		Number of cells & 21\\
		$R/L$ [M$\Omega /$m]& 13.9\\
		Aperture [mm] & 40\\
		Mode separation (in $\pi$ mode) [MHz] & 5.3 \\
		RF Pulse length [\SI{}{\us}] & 3 \\
		Coupling factor & 2 \\
\end{tabular}
\end{ruledtabular}
\end{table}

\subsection{Solenoids Around RF Cavities}
\label{ch:sols}

\begin{figure}[h!]
    \centering
    \includegraphics[width=0.45\textwidth]{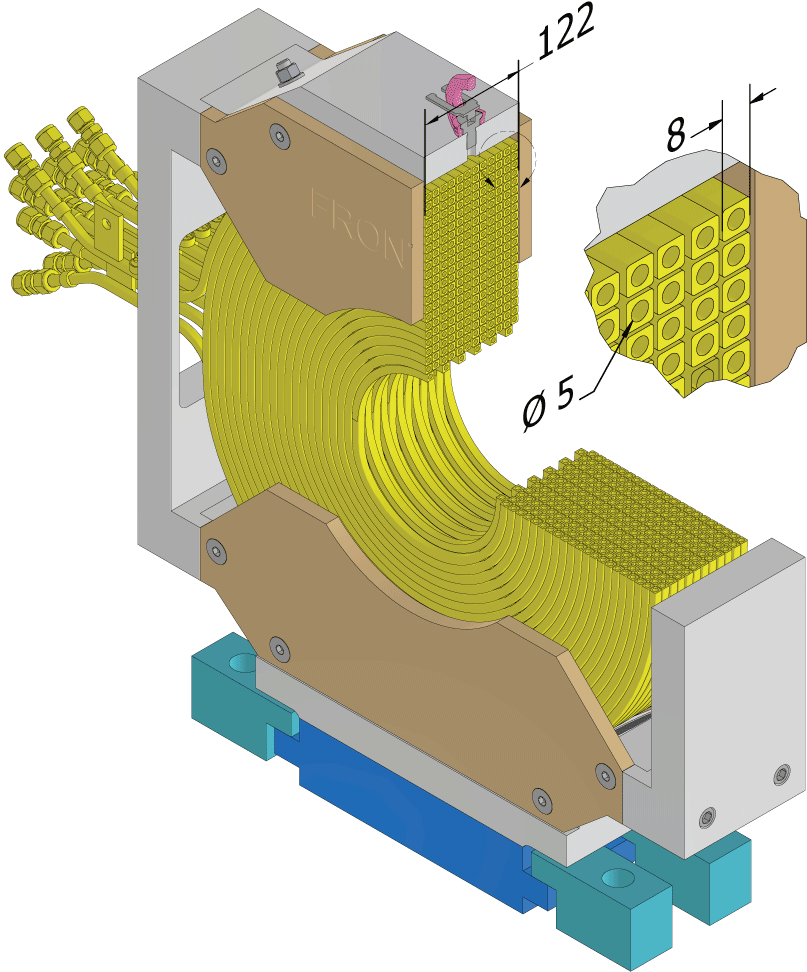}
    \caption{\label{fig:solenoids} Mechanical design of solenoids around RF cavities and supporting structure. Including section and detail view of windings and cooling channels.}
\end{figure}

16 normal conducting solenoids will surround the RF cavities, almost uniformly distributed along the capture section as shown in Fig.~\ref{fig:P3Layout}. Each of whom will generate a peak field of 0.22~T, and combined they will deliver the desired 0.45~T plateau along the beam axis. These solenoids, shown in Fig.~\ref{fig:solenoids}, are winded into 22 layers and 12 helical windings per layer, with 5~mm diameter cooling water channels. The device length is 112~mm, and the aperture and outer diameter 160~mm and 556~mm respectively. Two arrangements of 8 solenoids (one per RF cavity, see layout in Fig.~\ref{fig:P3Layout}) will be fed separately with 220~A, equivalent to a current density limit of 5~A/mm$^2$. No iron yoke is considered for maximum field flatness. The voltage drop per solenoid is 30 V, which results in a power consumption of 6.5~kW. Cooling can be provided through six channels, with a maximum pressure and water speed of 1~bar and 1~m/s, as well as an inlet-outlet temperature raise of \SI{20}{\celsius}. Each solenoid will be encased in an Aluminium support, as illustrated in Fig.~\ref{fig:solenoids}, that will withstand the individual 130~kg copper weight as well as the forces exerted by other solenoids. According to simulations, such forces would be as high as 23.5~kN, reaching the peak in the most upstream normal conducting solenoid due to its proximity to the HTS peak field. Notice also in Fig.~\ref{fig:HTScoils} the large supporting structure around the HTS cryostat. 

\begin{table}[h]
\caption{Overview of normal conducting solenoid parameters}
\label{tab:RFCavitiesParams}
\begin{ruledtabular}
\begin{tabular}{lc}
		Length [mm] & 112\\
    	Coil diameter [mm] & 160 (inner), 556 (outer)\\
        Peak Field Single Solenoid [T] & 0.213\\
		Current [A] & 220\\
		Layers, Windings per layer & 22, 12\\
        Power Consumption per solenoid [kW] & 6.5
\end{tabular}
\end{ruledtabular}
\end{table}

\section{Beam Dynamics}
\label{ch:BD}

The RF and solenoid systems described in section~\ref{ch:KeyTech} will drive the beam with unprecedented efficiency from the target up to the experiment diagnostics. The beam dynamics associated to this capture system will be elucidated below, with a particular emphasis on few key factors behind the e+ yield upgrade: an abundant e+ production at the target, solenoid fields strong around the target and uniform along the RF cavities combined with a large iris aperture and a comprehensive RF phase optimization. Studies are supported by Geant4~\cite{G4} and ASTRA~\cite{ASTRA} simulations. 

\subsection{e+ Production at Target}
\label{ch:pProd}

\begin{figure*}
    \centering
    \includegraphics[width=0.9\textwidth]{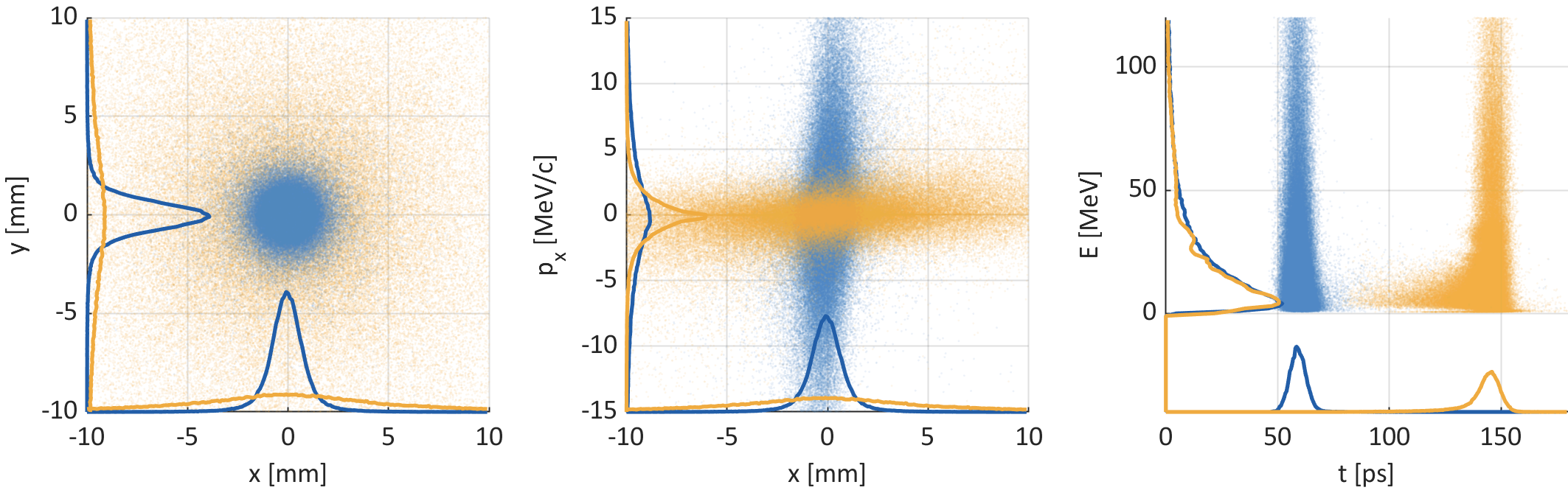}
    \caption{\label{fig:AMD_BD} Transverse and longitudinal profiles of the e+ distribution at exit face of the target (blue) and at the entrance of 1$^{st}$ RF cavity (yellow). Corresponding statistical values found in Table.~\ref{tab:TargetVSAMD}.}
\end{figure*}

200~pC e- at 6~GeV will impinge upon the 17~mm-thick Tungsten target according to parameters in Table~\ref{tab:FCCvsP3}, yielding a e+e- beam in the multi-MeV and nano-Coulomb range. Transverse and longitudinal profiles of the secondary e+ distribution are shown in Fig.~\ref{fig:AMD_BD}, and the corresponding list of parameters in Table~\ref{tab:TargetVSAMD}. Longitudinally, secondary bunches will have a length ($\sigma_t$~=~5.7~ps) comparable to that of primary e- ($\sigma_t$~=~3.3~ps). However, their uncorrelated energy spread ($\Delta$E$_{RMS}$~=~122.8~MeV) will be significantly greater. Similarly in the transverse plane, despite the moderate beam size at the target exit ($\sigma_x$~=~1.1~mm), e+ will have a large uncorrelated spread of transverse momentum ($\sigma_{px}$~=~7.1~MeV/c). These values indicate, as will be evidenced in the following paragraphs, that e+ dynamics are heavily dominated by extremely large energy spread and transverse emittance. 

\begin{table}[h]
\caption{Main parameters of charge, and transverse and longitudinal dynamics of the e+ beam at the target exit and the entrance of the 1$^{st}$ RF cavity. }
\label{tab:TargetVSAMD}
\begin{ruledtabular}
\begin{tabular}{lcc}
		& Target exit\footnote{Simulated with Geant4~\cite{G4} according to parameters in Table~\ref{tab:FCCvsP3}}  & 1$^{st}$ RF cav. entrance\footnote{Simulated with ASTRA~\cite{ASTRA} on the basis of $^a$}\\
        \colrule
        $Q_{e+}$ [pC]                          & 2754      & 2334\\
  	     $N_{e+}/N_{e-}$                          & 13.77      & 11.67\\
        \colrule
        $\sigma_x$, $\sigma_y$ [mm] & 1.1 & 6.2 \\
		$\sigma_{px}$, $\sigma_{py}$  [MeV/c] & 7.1 & 2.7\\
        $\epsilon_{x,norm}$, $\epsilon_{y,norm}$ [\textit{$\pi$ mm mrad}]& 11676 & 12016\\
        \colrule
        $\sigma_t$ [ps] & 5.7 & 11.3\\
  	$\Delta$E$_{RMS}$ [MeV] & \multicolumn{2}{c}{122.8}\\
\end{tabular}
\end{ruledtabular}
\end{table}

\begin{figure}[h!]
    \centering
    \includegraphics[width=0.45\textwidth]{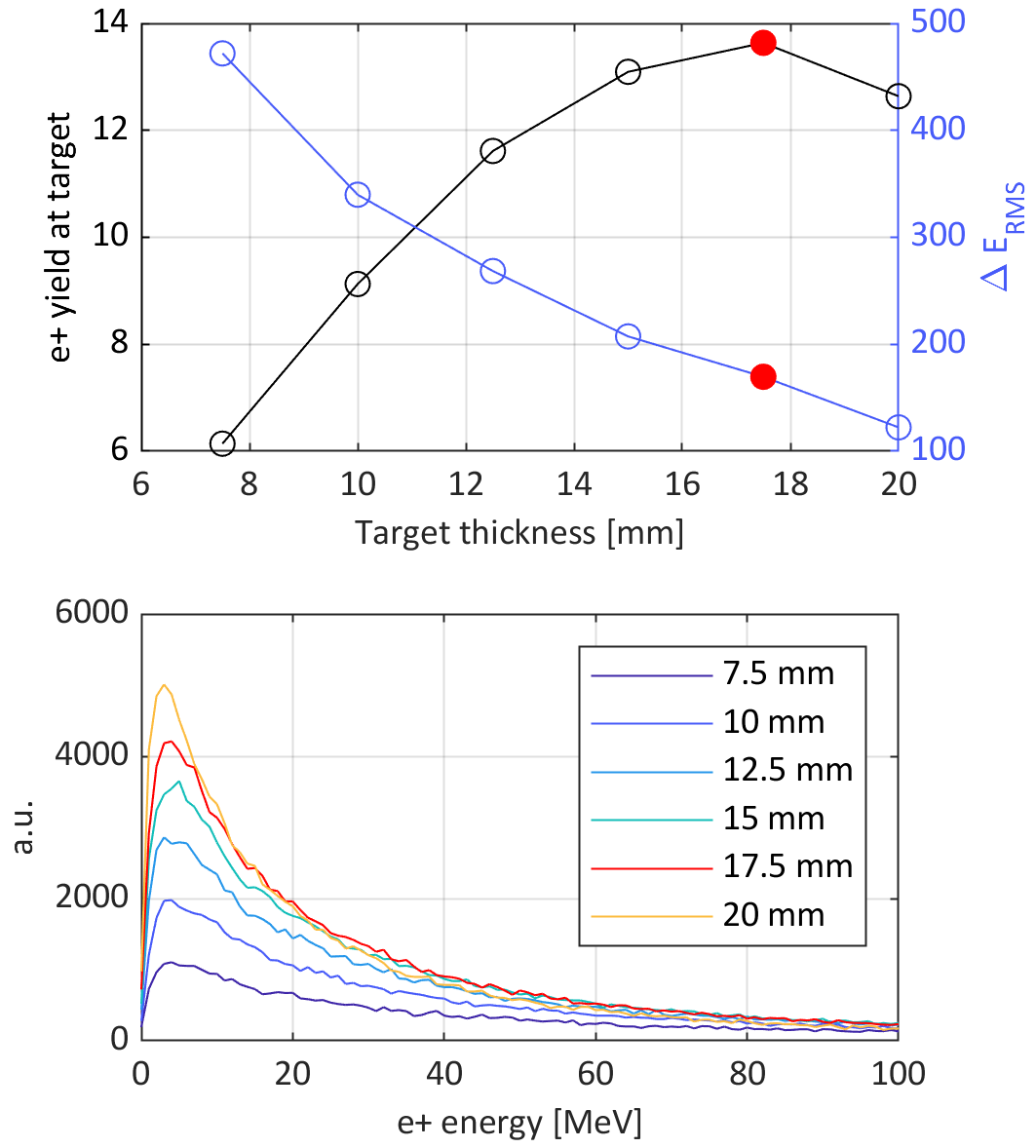}
    \caption{\label{fig:TargetLong} e+ yield at the target and RMS energy spread with respect to target thickness, assuming 6~GeV primary e- beam (top). e+ energy distributions for different target thickness values (bottom). Current baseline, 17.5~mm, shown in red.}
\end{figure}

Primary e- beam parameters are inherited from the FCC-ee baseline ---with the exception of beam current--- which are the result of previous optimization works~\cite{Chaikovska:FCC-ee}. The rule of thumb for target-based e+ sources is that higher primary e- energies and smaller transverse beam sizes will provide greater yields. In addition, there is an optimum target thickness for each given primary energy value~\cite{Zhao:CLICOpt}, which in the FCC-ee case is 17.5~mm. A low-key benchmarking and optimization study was performed for P$^3$, obtaining the same results as above. Fig.~\ref{fig:TargetLong} shows a target thickness scan with 6~GeV, 0.5~mm impinging e-, where a clear maximum (13.77~$N_{e+}/N_{e-}$) is reached also at 17.5~mm. It is also observed that a greater target thickness can reduce the RMS energy spread, but the 17.5~mm baseline is maintained. 

\begin{figure}[h!]
    \centering
    \includegraphics[width=0.45\textwidth]{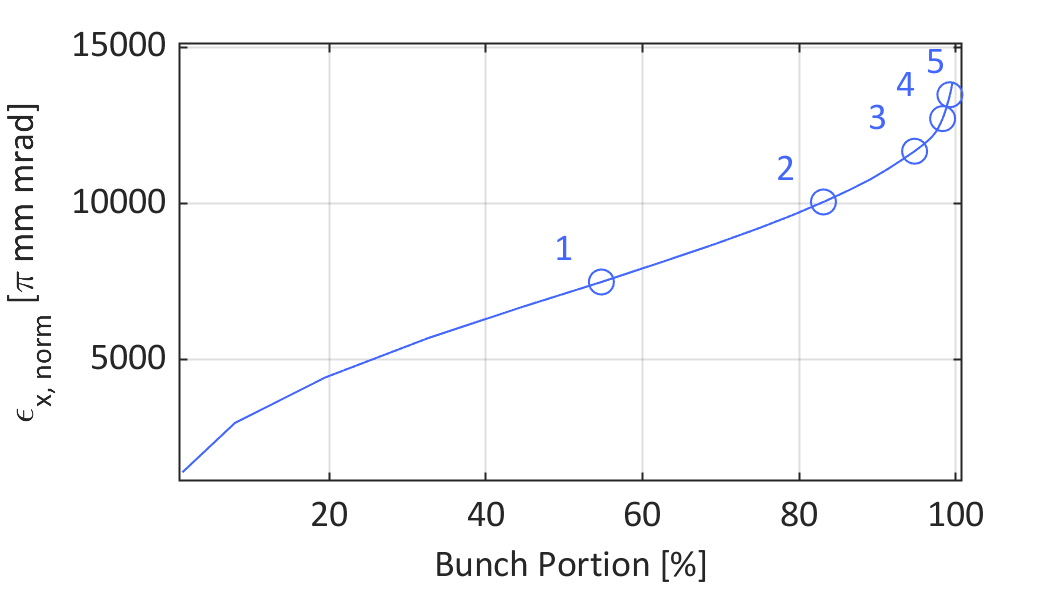}
    \caption{\label{fig:SliceEmit} Normalized transverse core e+ emittance~\cite{floettmann:emit} at the target exit for different beam slices, defined as Twiss ellipse fittings. Numbers next to data points represent $\sigma$, or the semi-axes sizes, 1$\sigma$ being the nominal Twiss parameters. }
\end{figure}

Due to the extreme transverse emittance inherent to pair production, some e+ may emerge from the target with an exceptionally high transverse momentum $p_x$, leading to divergence values ($p_x$/$p_z$) well above~1. Despite representing a small portion of the e+ charge, these particles may significantly inflate the computed emittance. To avoid these effects, core emittance~\cite{floettmann:emit} is calculated for different beam slices, defined within the equivalent twiss ellipses. As shown in Fig.~\ref{fig:SliceEmit}, particles encompassed by the nominal Twiss ellipse (1$\sigma$) have an emittance of 7486~\textit{$\pi$ mm mrad}. The 3$\sigma$ ellipse, which comprises 95~$\%$ of e+ at the target exit, yields an emittance of 11676~\textit{$\pi$ mm mrad}. In this framework, there is little margin for emittance reduction, since secondary $p_x$ spread is fairly insensitive to primary e- energy and size. However, further e+ yield maximization and emittance cooling through alternative target geometries are currently under study and will be tested during the P$^3$ experiment. 

\subsection{Transverse e+ Capture through High Solenoid Fields}
\label{ch:TransBD}

Like most preceding e+ linac designs~\cite{Chaikovska:PositronSources}, P$^3$ relies on a solenoid system for e+ collection, aiming to transport extreme emittances with the highest possible efficiency up to the DR where e+ will be cooled. To this end, the solenoid arrangement and field profile shown in Fig.\ref{fig:P3Layout} make up an adiabatic matching device (AMD)~\cite{Helm:AMD, chehab:CAS2}, a well-known e+ capture technique based on transforming the transverse phase space of newly generated e+ (moderate $\sigma_x$ and large $\sigma_{px}$) into the acceptance of the capture system (large $\sigma_x$ and moderate $\sigma_{px}$). This is obtained through high peak solenoid fields around the target (12.7~T), slowly decreasing towards a weaker magnetic plateau around the capture line (0.45~T).  The effect of such AMD is clearly illustrated in Fig.~\ref{fig:AMD_BD} and listed in Table~\ref{tab:TargetVSAMD}, providing great compression of the transverse momentum spread ($\sigma_{px}$~=~2.7~MeV/c) compensated by a beam size growth ($\sigma_{x}$~=~6.2~mm) well below the aperture of the RF cavities (40~mm diameter). 

\begin{figure}[h!]
    \centering
    \includegraphics[width=0.45\textwidth]{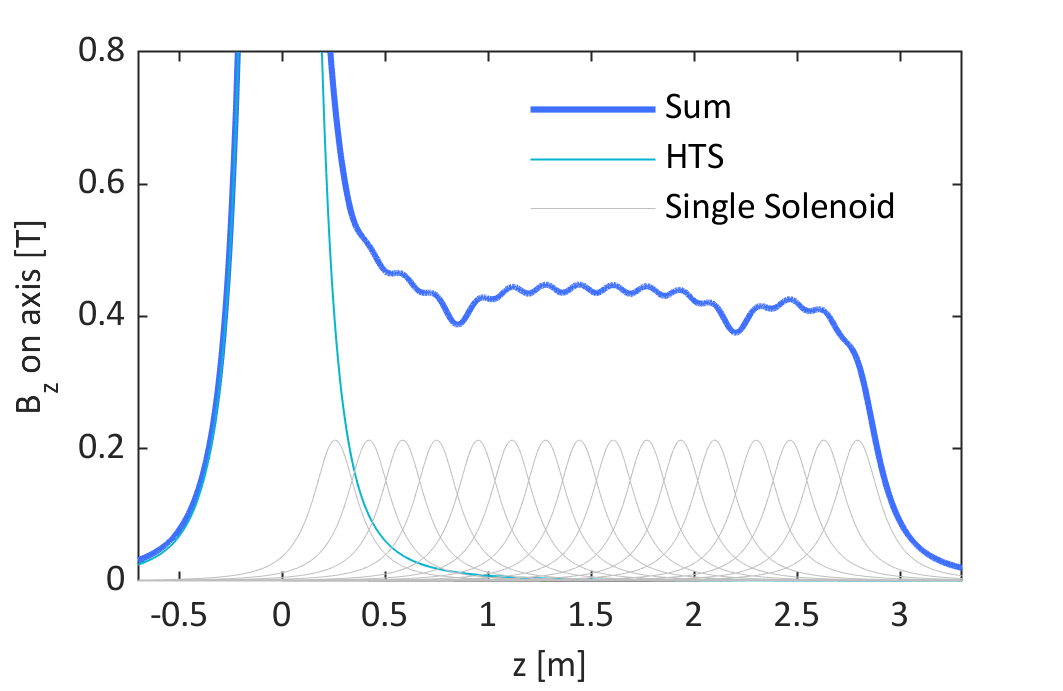}
    \caption{\label{fig:solenoidPlateau} Detail view of solenoid field profile on axis along the capture section. Including contributions of HTS and normal conducting solenoids.}
\end{figure}

\begin{figure}[h!]
    \centering
            \includegraphics[width=0.45\textwidth]{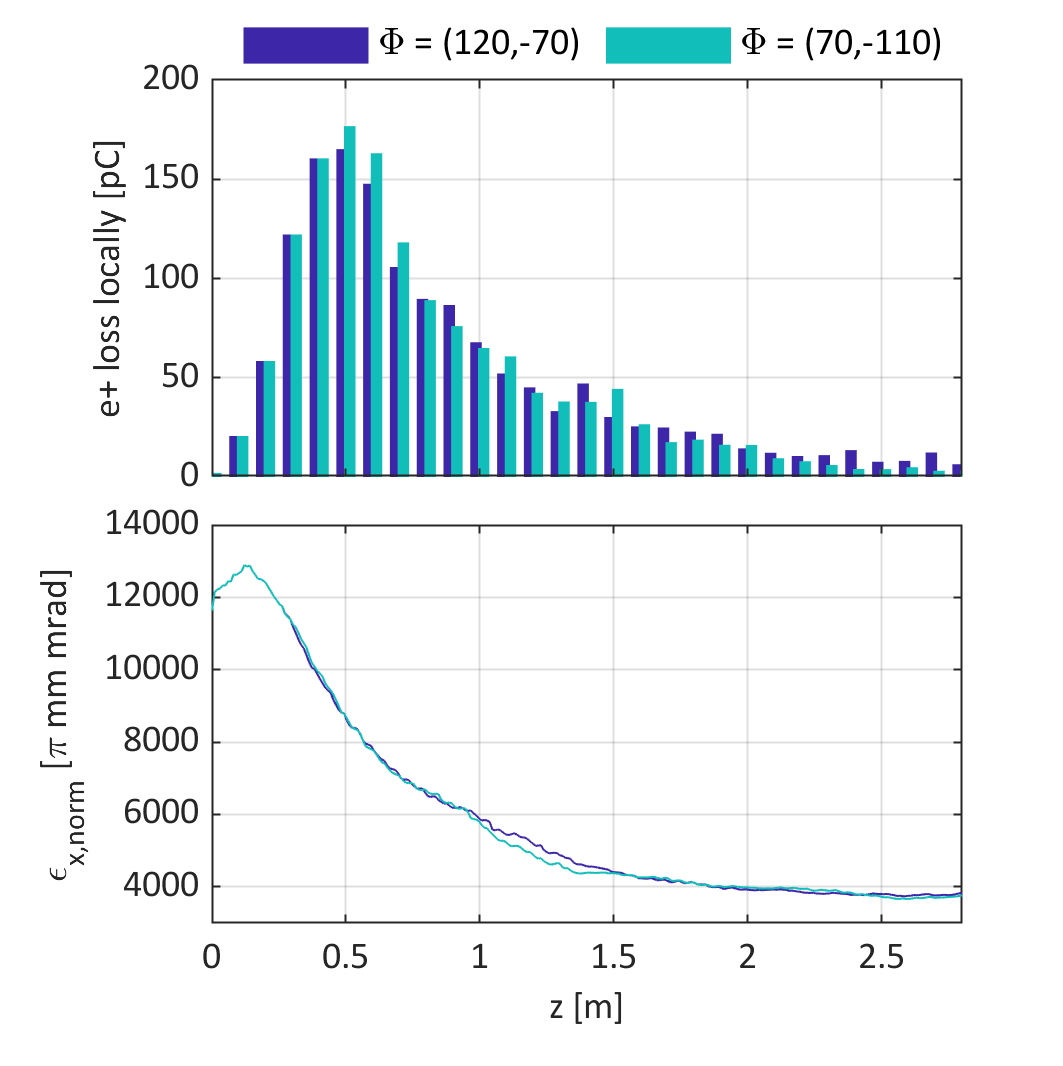} 
    \caption{\label{fig:EmitAndEff} Simulated local e+ charge loss (top) and normalized transverse emittance (bottom) along the RF cavities. Calculated for for RF working points of interest calculated in section~\ref{longBD}.}
\end{figure}

The 0.45~T plateau delivered by the NC solenoids will create a magnetic channel along the cavities with the ability to capture and transport a large proportion of the matched e+. A fairly uniform magnetic profile is observable in Fig.~\ref{fig:solenoidPlateau} besides relatively small drops around z~=~0.86~m and z~=~2.21~m, where the separation between solenoids is slightly incremented in order to fit the central waveguide couplers of the RF cavities. According to Fig.~\ref{fig:EmitAndEff}, the proposed solenoid arrangement would provide capture efficiencies as high as 45.7$\%$ with the ability to transport transverse normalized emittances around 4000~$\pi\ mm\ mrad$. Higher transmission rates are reachable through the use of multi-Tesla magnetic channels generated by low temperature superconducting solenoids, as considered in previous versions of the P$^3$ capture line design~\cite{Vallis:Linac22}. However, 0.45~T fields show a good performance while avoiding excessive costs and power consumption.

\subsection{RF Bunching and Acceleration}
\label{longBD}

\begin{figure}[h!]
    \centering
        \begin{subfigure}[b]{0.45\textwidth}
            \includegraphics[width=\textwidth]{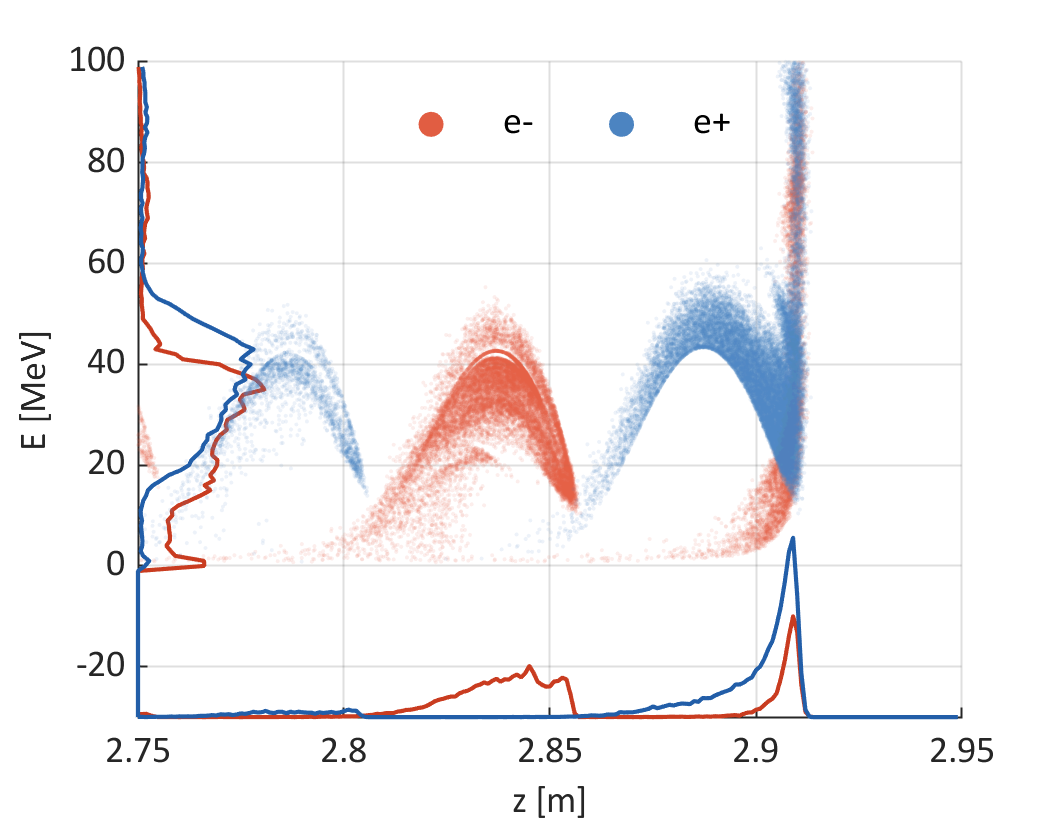}
            \caption{\label{fig:WP_a}}
        \end{subfigure}
        \begin{subfigure}[b]{0.45\textwidth}
            \includegraphics[width=\textwidth]{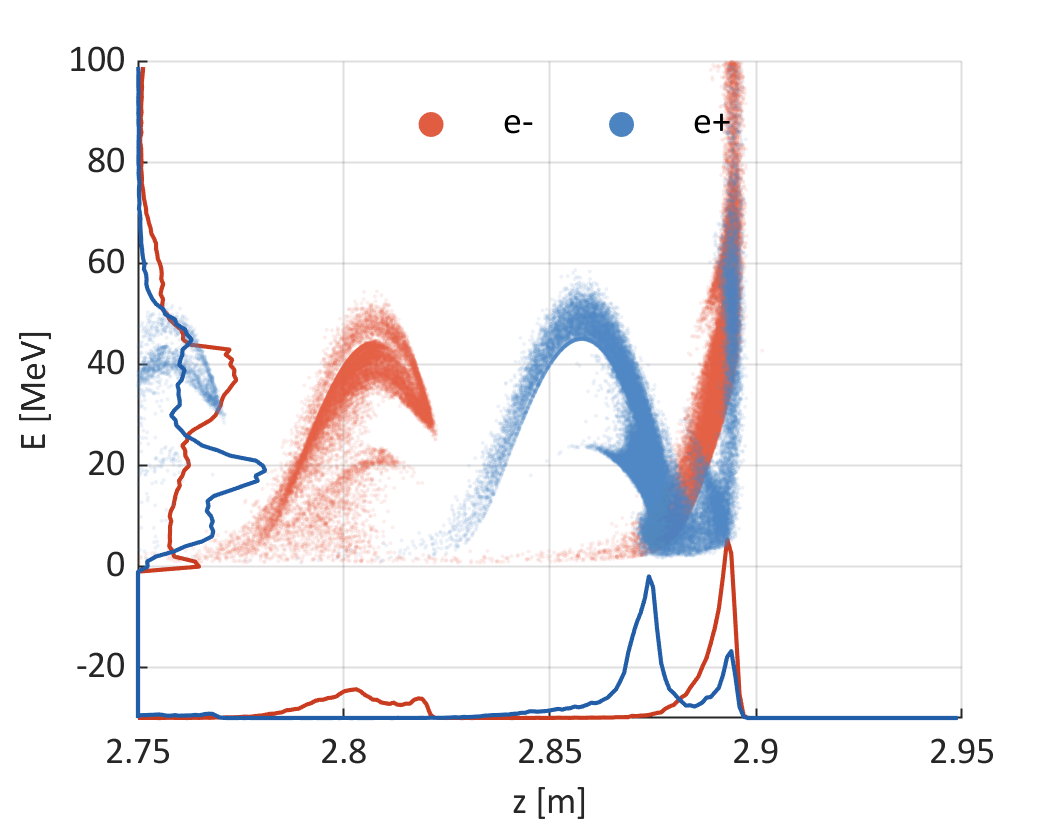}
            \caption{\label{fig:WP_b}}
        \end{subfigure}   
    \caption{\label{fig:WPs} Simulated e+e- distributions near the exit of the 2$^{nd}$ RF cavity (z~$\approx$~2.8~m) for RF working points of interest: $\Phi$~=~(120,-70) (a) and $\Phi$~=~(70,-110) (b). Notice that  longitudinal dimension is given in units of length.}
\end{figure}

Secondary e+e- will emerge from target under the influence of the 12.7~T solenoid field, describing spiralling trajectories with a wide wange of Larmor angles and \textit{radii} due to their extremely large energy spread~\cite{chehab:CAS}. This effect will make newly generated particle distributions grow longitudinally ($\sigma_t$~=~11.3~ps), as illustrated in Fig.~\ref{fig:AMD_BD} and Table.~\ref{tab:TargetVSAMD}. As a result, RF fields will generate time structures of many consecutive e+ and e- bunches separated by $\lambda$/2 (167~ps or 50~mm in the ultrarelativistic regime), particularly populated in the first two RF buckets. The typical bunching profile of P$^3$, despite being highly dependent on the RF phase configuration, is well depicted in Fig.~\ref{fig:WPs}.

Regarding the RF optimization, two main figures of merit (FOMs) are considered: the total captured e+ charge at the exit of the 2$^{nd}$ RF cavity and the equivalent e+ yield at the FCC-ee DR (see section~\ref{ch:Intro}). While the first FOM corresponds to a real, measurable quantity, the latter will establish a correction factor with respect to the equivalent charge accepted at the FCC-ee DR. Such yield is computed through e+ tracking up to 200~MeV, extending the baseline simulation layout from 2 to 10 RF cavities surrounded by solenoids. The resulting longitudinal time-energy distribution at 200~MeV is transformed analytically up to 1.54~GeV, the nominal energy of the FCC-ee DR, where a filter in energy of $\pm$3.8$\%$ is applied. Notice that this analytical approach is proven highly accurate with respect to 6D particle tracking simulations, due to the small transverse losses above 200~MeV and weak radial dependency of the accelerating E-field within the iris area~\cite{Zhao:AMDOptimization}. 

\begin{figure}[h!]
    \centering
        \begin{subfigure}[b]{0.45\textwidth}
            \includegraphics[width=\textwidth]{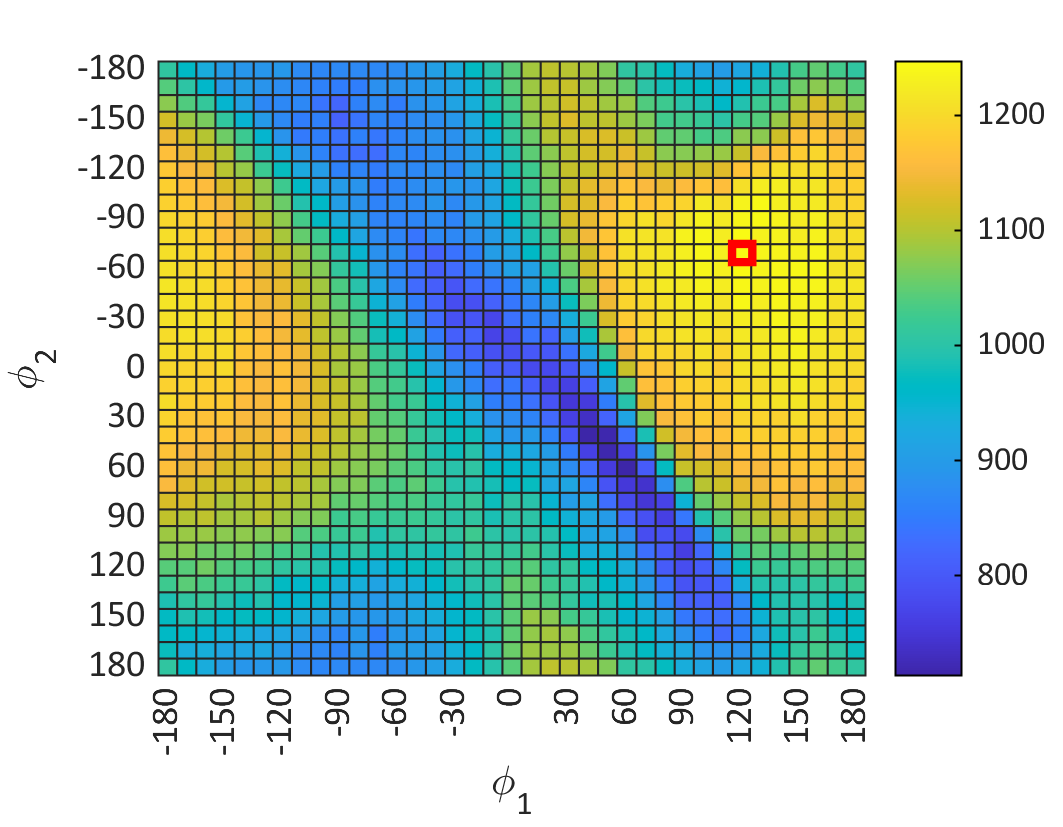}
           
            \caption{ \label{fig:2DScan_a}}
        \end{subfigure}
        \begin{subfigure}[b]{0.45\textwidth}
            \includegraphics[width=\textwidth]{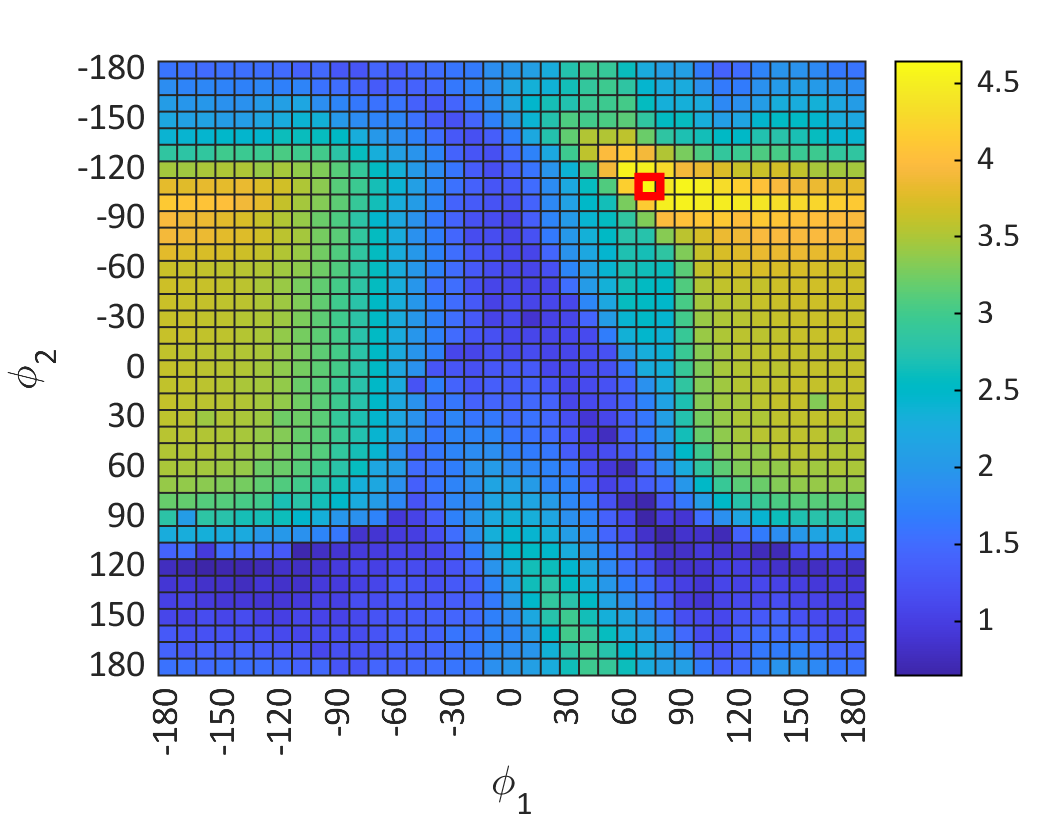}
            
            \caption{\label{fig:2DScan_b}}
        \end{subfigure}  
    \caption{\label{fig:2DScan} Total e+ captured charge in pC (a) and estimated e+ yield at the FCC-ee DR (b) simulated over full 2D RF phase scan. $\Phi=$~(120,-70) and $\Phi=$~(70,-110) marked in red in (a) and (b) respectively.}
\end{figure}

Both figures of merit, as represented in Fig.~\ref{fig:2DScan}, are strongly determined by the RF phase of both cavities. Among all possible RF configurations, 2 working points of interest (see Table~\ref{tab:Yields}) were chosen: $\Phi=$~(120,-70), which provides maximum e+ capture of 1246~pC after the 2$^{nd}$ RF cavity and $\Phi=$~(70,-110), corresponding to the maximum e+ Yield of 4.64~$N_{e+}/N_{e-}$ at the FCC-ee DR. Notice that due to the large beam spread, it is difficult to agree upon conventions such as the bunch center. For this reason, the RF phases introduced throughout this paper are arbitrary, with notions such as crest or zero-crossing having no particular physical meaning.

\begin{table}[h]
\caption{e+ charge and yield provided by RF working points of interest.}
\label{tab:Yields}
\begin{ruledtabular}
\begin{tabular}{lcc}
		& 2$^{nd}$ RF cav. exit  & FCC-ee DR\\
        \colrule
        \multirow{2}{*}{$\Phi$~=~(120,-70)} & 1246 pC  & 768 pC\\
                                                                          &   6.23~$N_{e+}/N_{e-}$ & 3.84~$N_{e+}/N_{e-}$\\
        \colrule
  	    \multirow{2}{*}{$\Phi$~=~(70,-110)} & 1153 pC  & 928 pC \\
                                                                           &  5.77~$N_{e+}/N_{e-}$ & 4.64~$N_{e+}/N_{e-}$\\
\end{tabular}
\end{ruledtabular}
\end{table}

\begin{figure}[h!]
    \centering
        \begin{subfigure}[b]{0.45\textwidth}
            \includegraphics[width=\textwidth]{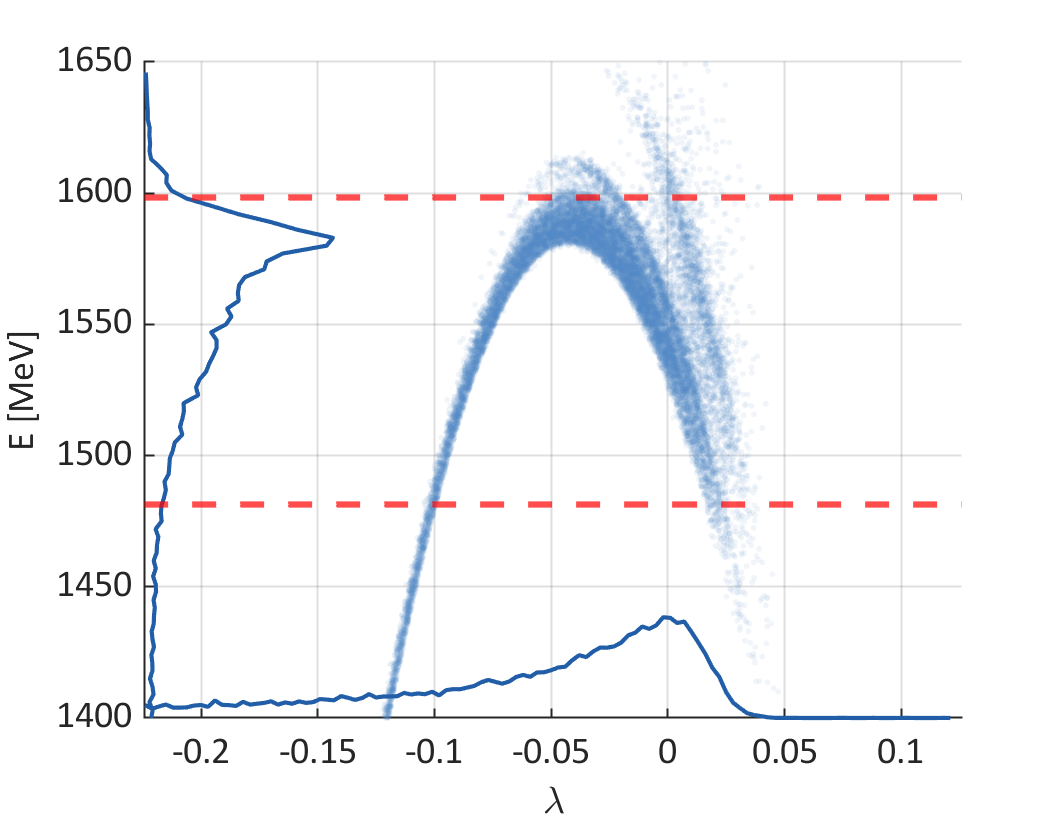}
            \caption{\label{fig:DR_WP_a}}
        \end{subfigure}
        \begin{subfigure}[b]{0.45\textwidth}
            \includegraphics[width=\textwidth]{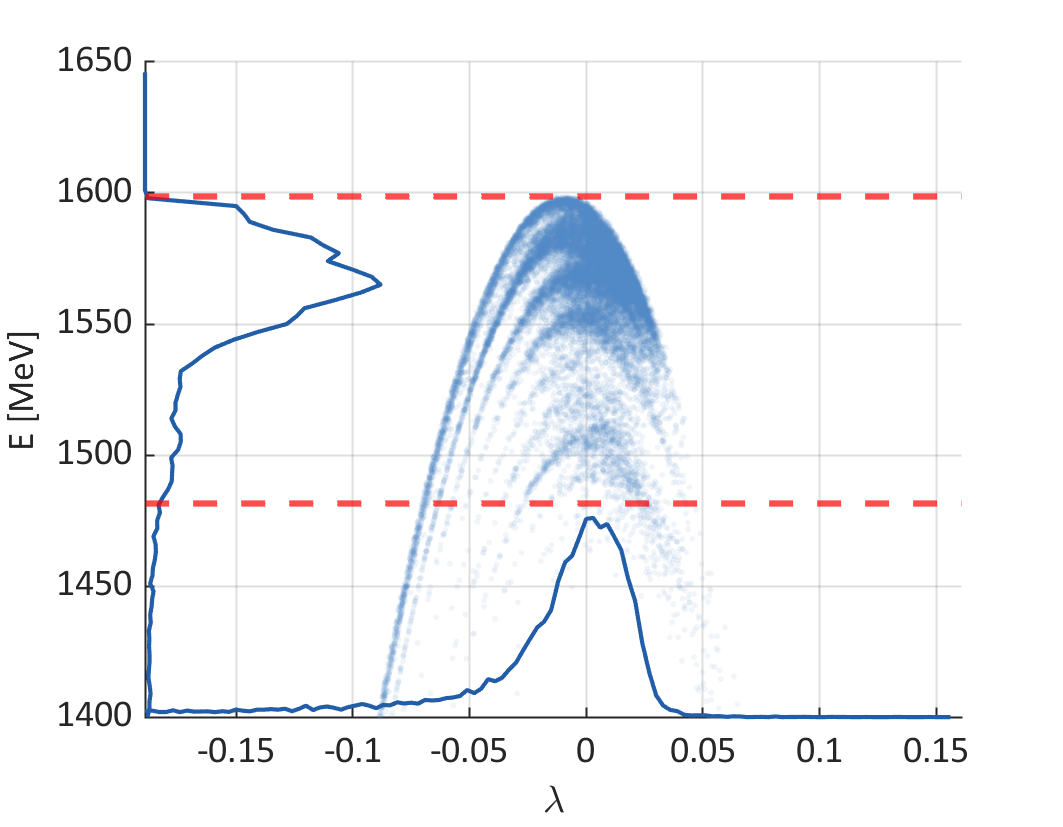}
            \caption{\label{fig:DR_WP_b}}
        \end{subfigure}
    \caption{\label{fig:DR_WP} Estimated e+ distributions at entrance of the FCC-ee DR for RF working points of interest: $\Phi$~=~(120,-70) (a) and $\Phi$~=~(70,-110) (b). Red dashed lines represent energy acceptance of DR, $\pm$3.8$\%$ of 1.54~GeV. Notice that  longitudinal dimension is given in terms of wavelength ($\lambda$), zero corresponding to the RF crest.}
\end{figure}

Major differences between the RF working points of interest can be observed at the exit of the 2$^{nd}$ RF cavity. Fig.~\ref{fig:WP_a}, corresponding to $\Phi$~=~(120,-70), shows an extremely spread and smooth energy profile ranging from below 20~MeV to above 50~MeV. Longitudinally, e+ are concentrated towards the high energy tail downstream from the main RF bucket. Instead, the $\Phi$~=~(70,-110) case illustrated in Fig.~\ref{fig:WP_b}, shows a greater e+ population towards the crest of the main RF bucket. In addition, despite also having a large spread in energy, a clear concentration peak is observed around 20~MeV. Such differences are even more noticeable at the DR entrance energy of 1.54~GeV. The excellent bunching provided by the $\Phi$~=~(70,-110) setup, illustrated in Fig.~\ref{fig:DR_WP_b}, explains such a high DR acceptance. It contrasts with the largely spread energy profile shown in Fig.~\ref{fig:DR_WP_a}, which leads a lower yield at the FCC-ee DR despite having the highest e+ charge capture rates. 

\begin{figure}[h!]
    \centering
            \includegraphics[width=0.45\textwidth]{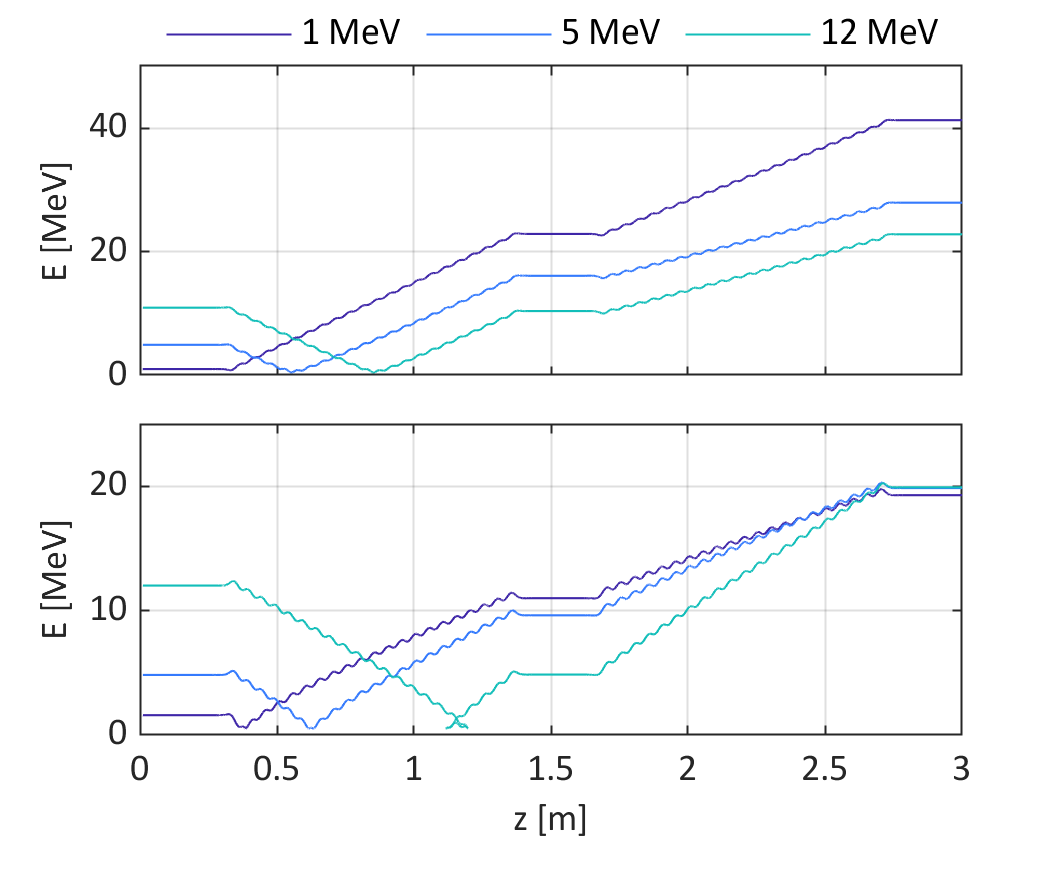} 
    \caption{\label{fig:Eref} Simulation of e+ at different initial energies accelerated by the P$^3$ RF cavities, for RF working points of interest: $\Phi$~=~(120,-70) (top) and $\Phi$~=~(70,-110) (bottom).}
\end{figure}

\begin{figure}[h!]
    \centering
    \includegraphics[width=0.45\textwidth]{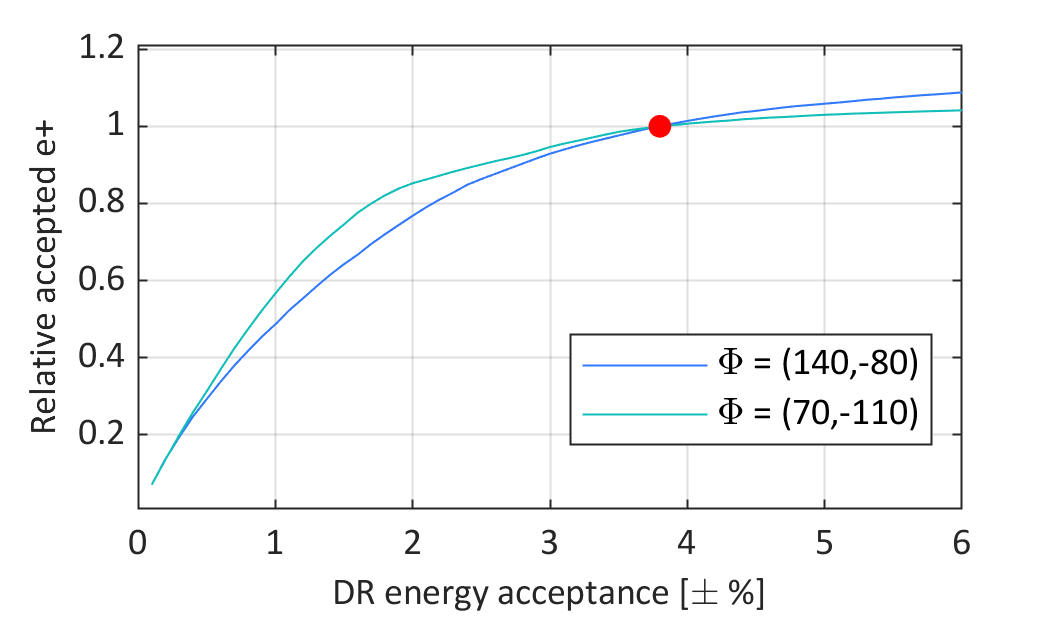}
    \caption{\label{fig:DRAcceptanceFactor} Relative yield at the DR with respect to the energy acceptance. Current baseline of $\pm$~3.8$\%$ is marked in red.}
\end{figure}

Fig.~\ref{fig:Eref} is particularly instructive about this energy compression process, showing how particles at 1~MeV, 5~MeV and 12~MeV reach almost the same energy at the exit of the 2$^{nd}$ RF cavity, achieved through a partial deceleration of the beam in the $\Phi$~=~(70,-110) case. Interestingly, the use of decelerating RF modes is a well-known energy compression technique particularly well suited for e+ sources~\cite{Aune-Miller:AMD}. Notice that the $\pm$~3.8~$\%$ energy acceptance filter of the DR may be subject to future changes. Thus, we provide in Fig.~\ref{fig:DRAcceptanceFactor} a scaling factor for different energies applied to both RF working points of interest for a range of DR acceptance parameters. 

\section{BEAM DIAGNOSTICS}
\begin{figure*}
    \centering
    \includegraphics[width=0.99\textwidth]{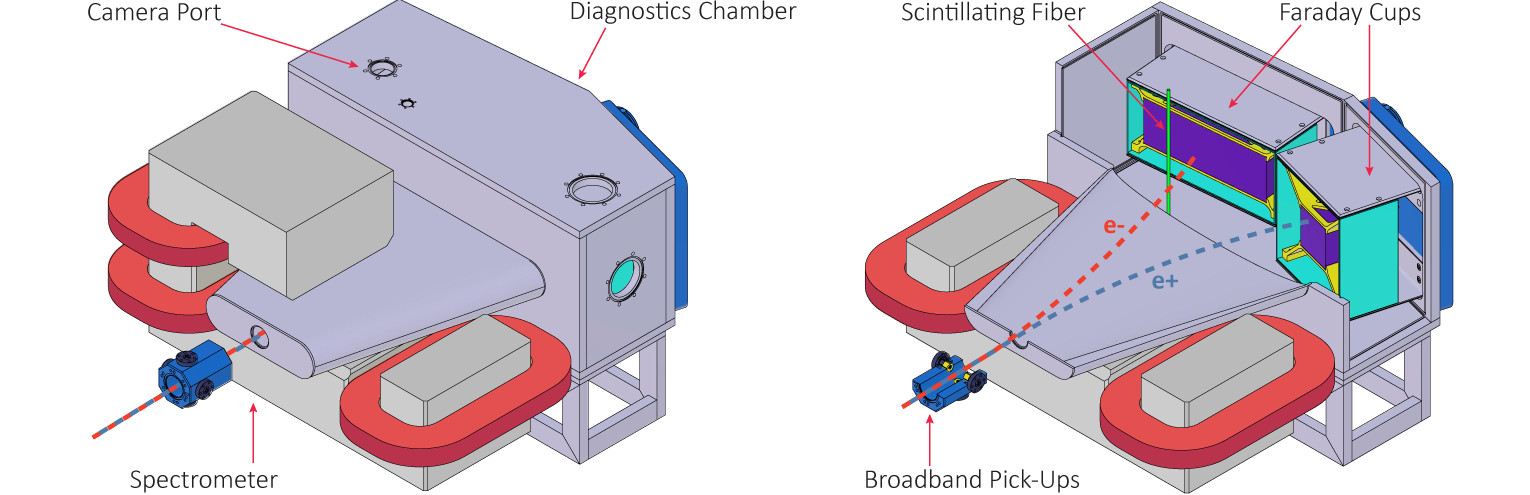}
    \caption{\label{fig:diagnostics} P$^3$ diagnostics setup (left) including inside view (right).}
\end{figure*}

A concept for the P$^3$ diagnostics is illustrated in Fig.~\ref{fig:diagnostics}. This setup will be equipped with an arrangement of broadband pick-ups (BBPs), 2 Faraday Cups (FCs)
and a variety of scintillating detectors. The BBPs will detect the time structure of the captured e+e- beam. The FCs and scintillators will be installed in the same vacuum chamber, and will measure the charge and energy spectrum of e+ and e- streams independently. Separation of particle species will be provided by a spectrometer, a dipole magnet based on four copper coils and an iron yoke, which will be fed at a maximum current of 340~A in order to reach magnetic fields up to 0.25~T.

\subsection{Broadband Pick-ups}

\begin{figure}[h!]
    \centering
    \includegraphics[width=0.45\textwidth]{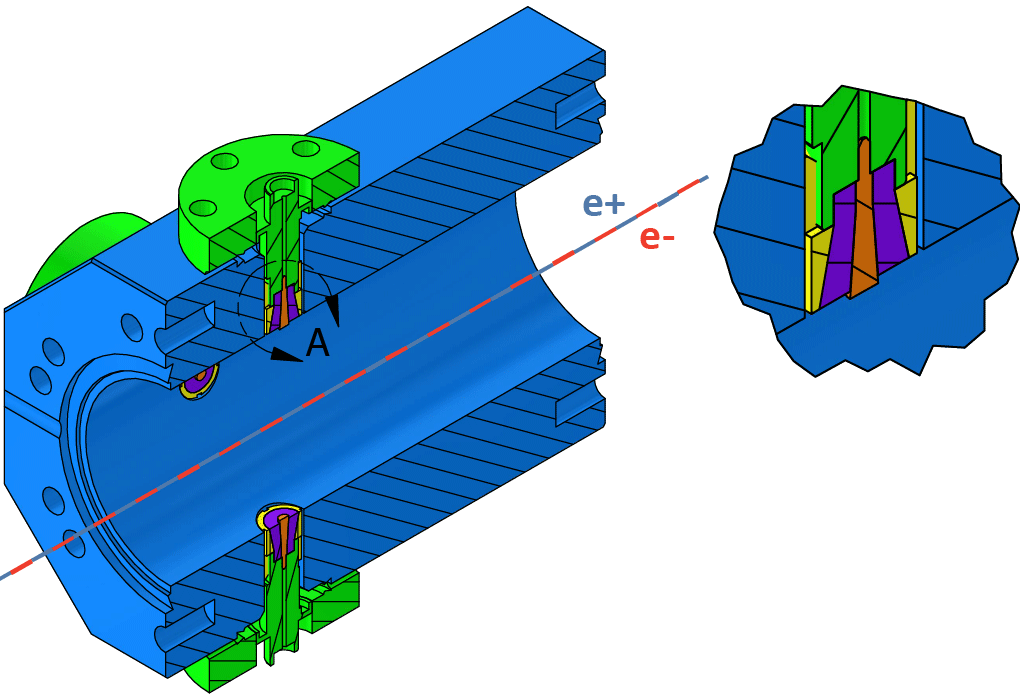}
    \caption{\label{fig:BBPs} Mechanical design of the 27~GHz BBP assembly. Detail view A features a pick-up (orange) with its corresponding dielectric PTFE holder (purple) and outer conductor (yellow) mounted on a feedthrough (green).}
\end{figure}

\begin{figure}[h!]
    \centering
    \includegraphics[width=0.45\textwidth]{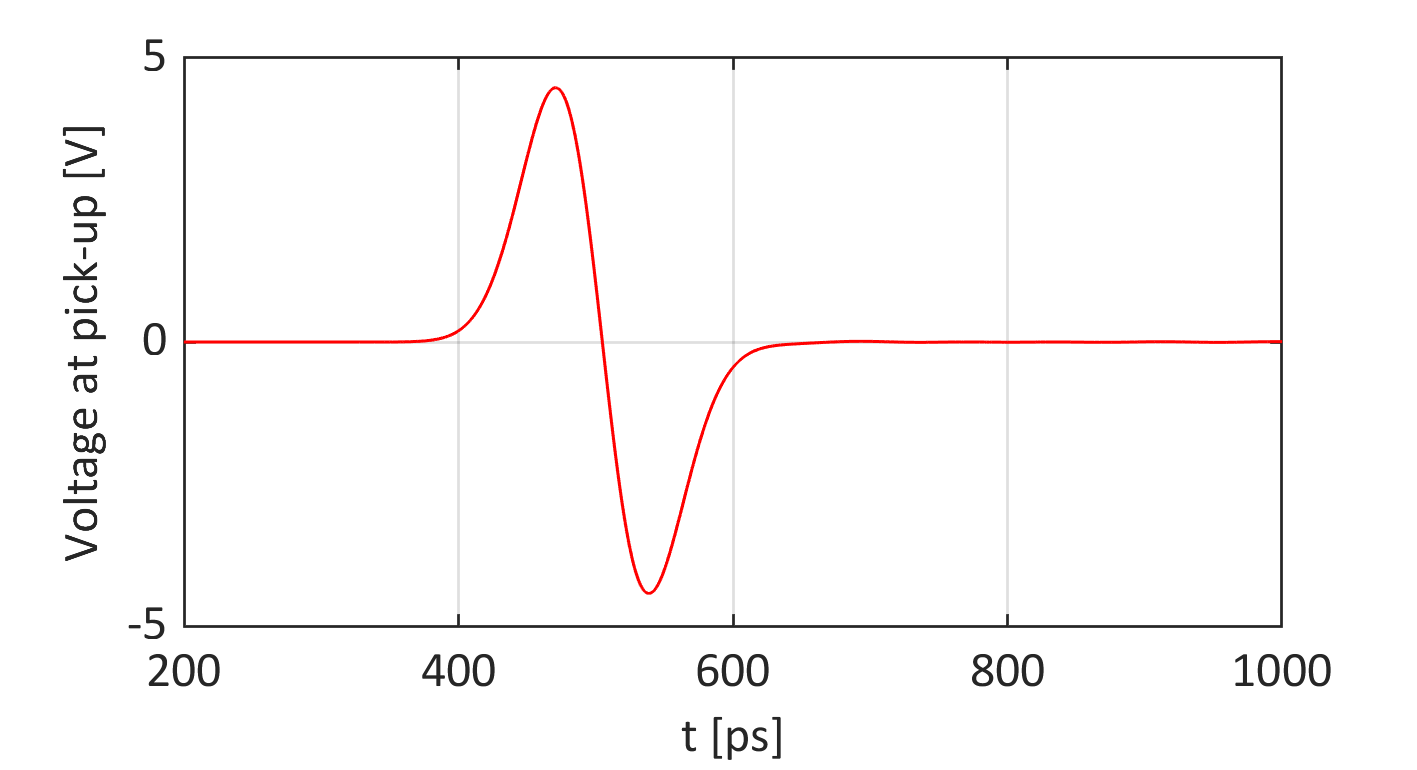}
    \caption{\label{fig:BBPs_signal} Detected voltage signal by one pick-up using 27~GHz feedthrough arrangement. Signal generated by gaussian bunch of Q$_{bunch}$~=~1~nC and $\sigma_t$~=~33~ps. Simulated with CST~\cite{CST}.}
\end{figure}

An arrangement of 4 broadband pick-ups (BBPs), shown in Fig.~\ref{fig:BBPs}, will follow the exit of the second RF cavity. The BBPs will detect simultaneously the wake voltage generated by the captured e+ and e- bunches in order to reconstruct the their time structure. This includes measurements of charge, length and separation bunch by bunch. The e+e- time structure will heavily depend on the RF phase, yet the typical distribution (see Fig.~\ref{fig:WPs}) will consist of alternating e+ and e- bunches of 33~ps length, and separated by 167~ps, namely half S-band period. Such fast measurements require an extremely broadband frequency response. Therefore, the geometry of the pick-ups was optimized to avoid intrinsic resonances up to frequencies in the range of a few tens of~GHz, while providing a relatively high peak voltage. According to a preliminary simulation based on a gaussian approximation of the P$^3$ bunches (see Fig.~\ref{fig:BBPs_signal}), the BBPs would detect a $\pm$4.5~V peak voltage signal with very small distortion. Notice that this simulation does not take into account major issues like wakefield effects, cable distortion or environment noise. 
 
Two BBP assemblies have been developed and fabricated, based on broadband feedthroughs of 27~GHz~\cite{Allectra27} (shown in Fig.~\ref{fig:BBPs}) and 65~GHz~\cite{Allectra65}. The hardware acquisition setup will consist of low attenuation, broad band cables~\cite{PasternackCables} and a high-end oscilloscope of at least 40~GHz pass band and 10~GS/s sampling. Both BBP assemblies are to be tested in different e- facilities before installation in P$^3$. These tests will not only allow to characterize and calibrate the BBPs, but also will elucidate important issues not covered by simulations, such as the performance of the BBPs under heavy noise sources and e+e- overlapping effects. Similar solutions based on ultra fast pick-ups are currently used in accelerator facilities such the fast BPMs at the SuperKEKB e+ linac~\cite{SuwadaBPM1,SuwadaBPM2} and the Bunch Arrival-Time Monitors at SwissFEL~\cite{SwissFELBAM}.

\subsection{Faraday Cups}

As illustrated in Fig.~\ref{fig:diagnostics}, the spectrometer will deflect the e+ and e- streams onto a highly asymmetrical arrangement of Faraday cups (FCs). Due to the extremely large energy (and p$_z$) spread, the main challenge for both FCs is to collect a beam with a remarkable transverse dispersion introduced at the spectrometer. As explained in the following paragraphs, each FC will follow an entirely different principle to address this issue. However, both of them would provide similar charge measurements of e+ and e-, which could be delivered to either FC through a sign inversion of the spectrometer polarity. 

\begin{figure}[h!]
\includegraphics[width=0.45\textwidth]{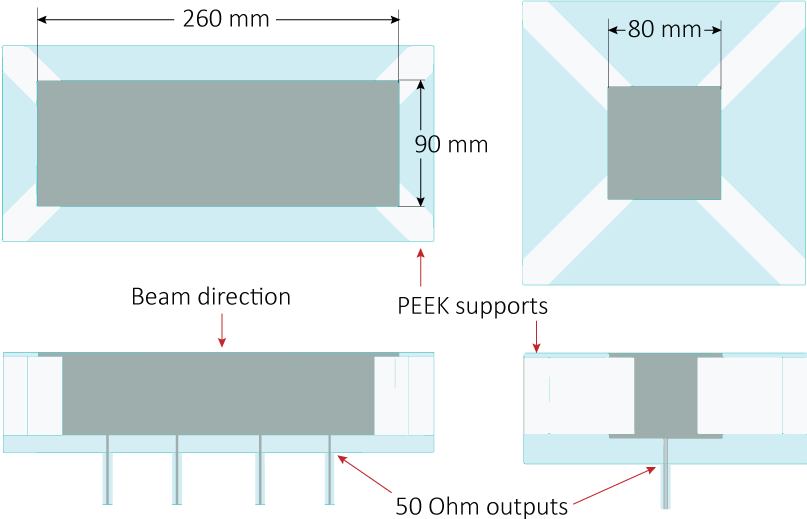}
\caption{\label{fig:FCLayout} Electromagnetic design of 12.5~$\Omega$ (left) and 50~$\Omega$ (right) FCs, including tungsten blocks, PEEK supports and vacuum space.}
\end{figure}

\begin{figure}[h!]
\includegraphics[width=0.45\textwidth]{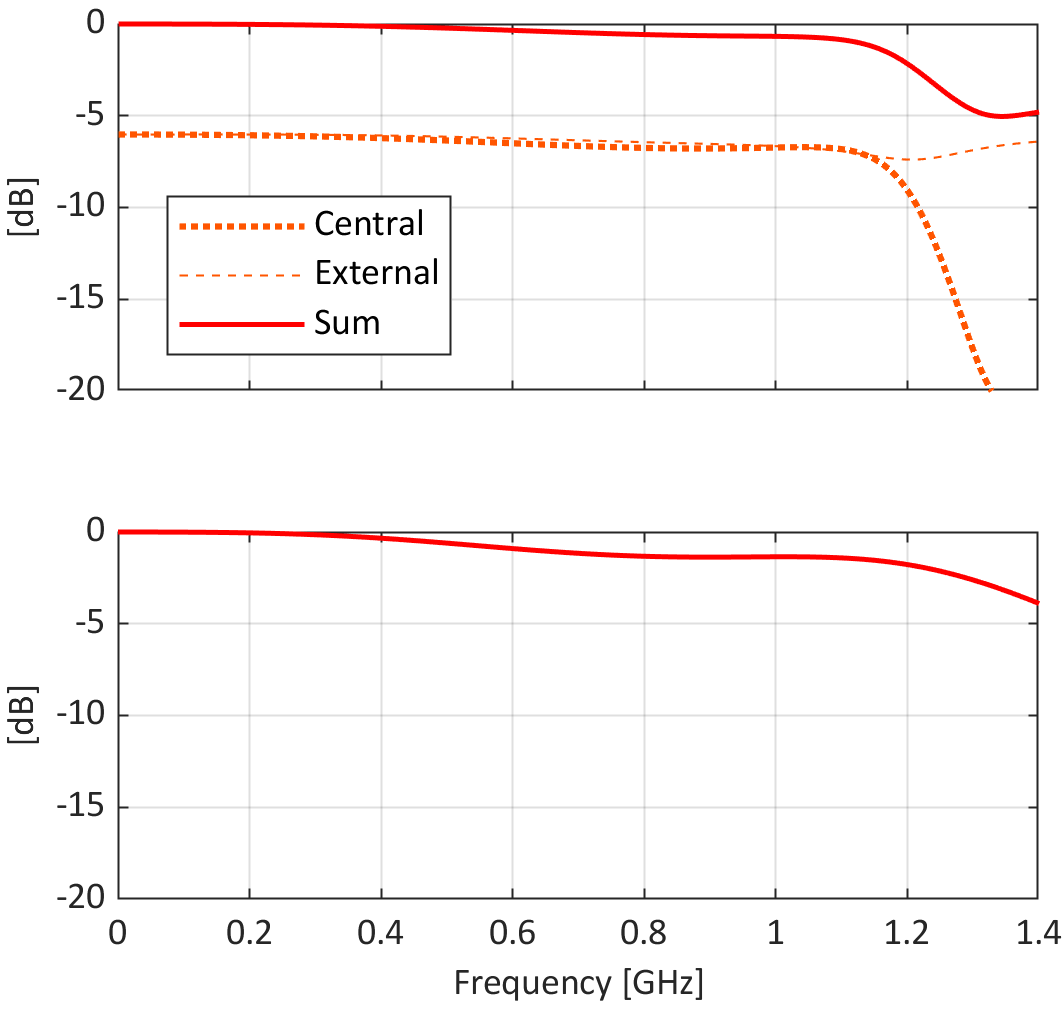}
\caption{\label{fig:FCResponse} Transmission parameters of 12.5~$\Omega$ (top) and 50~$\Omega$ (bottom) FCs. In the 12.5~$\Omega$ case, individual output  response included for innermost and outermost connectors. Based on HFSS~\cite{HFSS} simulations including tungsten blocks, PEEK supports and vacuum space, according to layouts illustrated in Fig.~\ref{fig:FCLayout}.}
\end{figure}

The first FC is tuned at 12.5~$\Omega$ in pursuit of a large transverse area (260x90~mm), which will maximize the collection of charged particles in a wide energy range of 9-75~MeV. The coaxial impedance, a factor 4 smaller than the 50~$\Omega$ standard, allows to reduce the size of the outer conductor and can be easily matched to standard circuits through the use of parallel 4 coaxial cables in the output, which will be read independently. A second, relatively compact FC (80x80~mm) tuned at 50~$\Omega$ will detect charged particles in a larger energy range of 3-90~MeV. Although the smaller transverse size does not allow for single-shot charge measurements in broad energy spectra, the 50~$\Omega$ FC will allow for energy discriminating measurements by adjusting the spectrometer strength, which determines the energy range of the particles routed towards the FC. A scan of 6 magnetic field values indicated in Table~\ref{tab:FCparams} would cover the above mentioned 3-90~MeV range. Fig.~\ref{fig:FCResponse} shows the frequency response of both FC arrangements with four diagonal PEEK supports (as seen in Fig.~\ref{fig:FCLayout}), in both cases above 1~GHz. 

\begin{table}[h]
\caption{Reference spectrometer strength and measured energy ranges for different channels of FCs. Values based on zero-emittance particles.}
\label{tab:FCparams}
\begin{ruledtabular}
\begin{tabular}{lcc}
	& Spectrom. strength [T] & Meas. E. range [MeV]\\
        \colrule
        12.5~$\Omega$ FC & 0.053 & 9 - 75 \\
        \colrule
        \multirow{5}{*}{50~$\Omega$ FC} & 0.212 & 50 - 90 \\
         & 0.120 & 28 - 50\\
	     & 0.068 & 16 - 28\\
         & 0.038 & 9 - 16\\
  	     & 0.021 & 5 - 9\\
         & 0.012 & 3 - 5
\end{tabular}
\end{ruledtabular}
\end{table}

\begin{figure}[h!]
        \begin{subfigure}[b]{0.45\textwidth}
            \includegraphics[width=\textwidth]{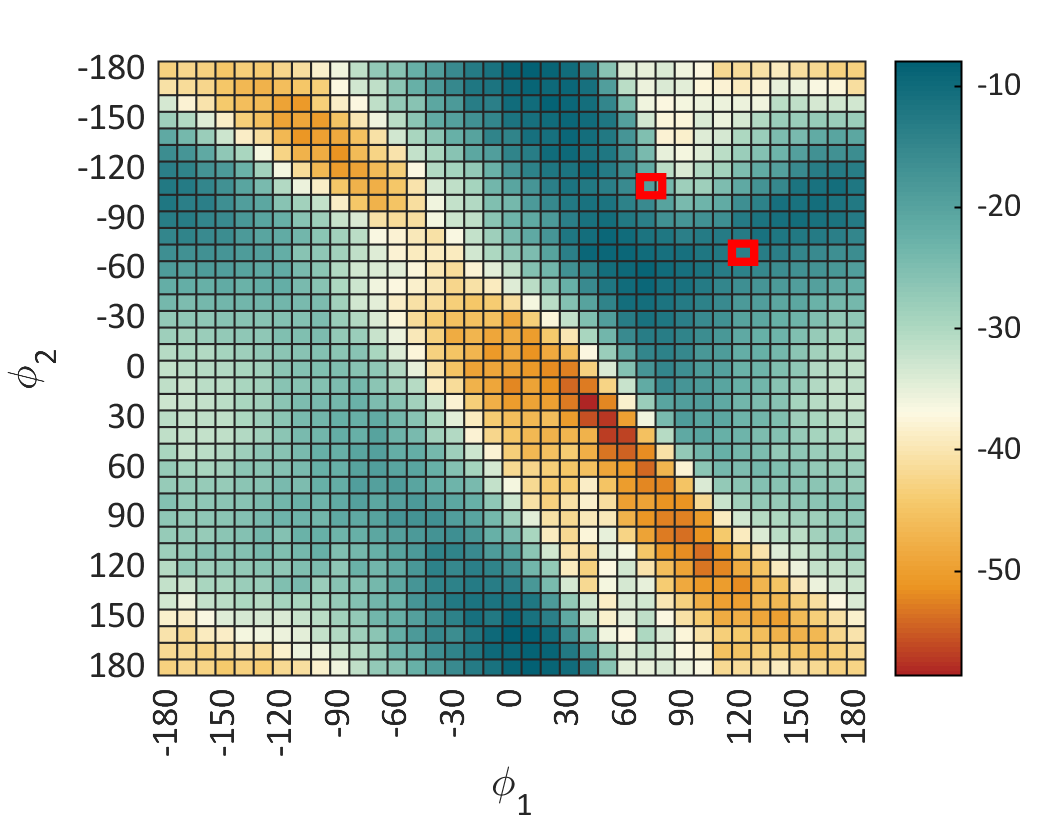}
            \label{fig:BBPs_a}
            \caption{}
        \end{subfigure}
        \begin{subfigure}[b]{0.45\textwidth}
            \includegraphics[width=\textwidth]{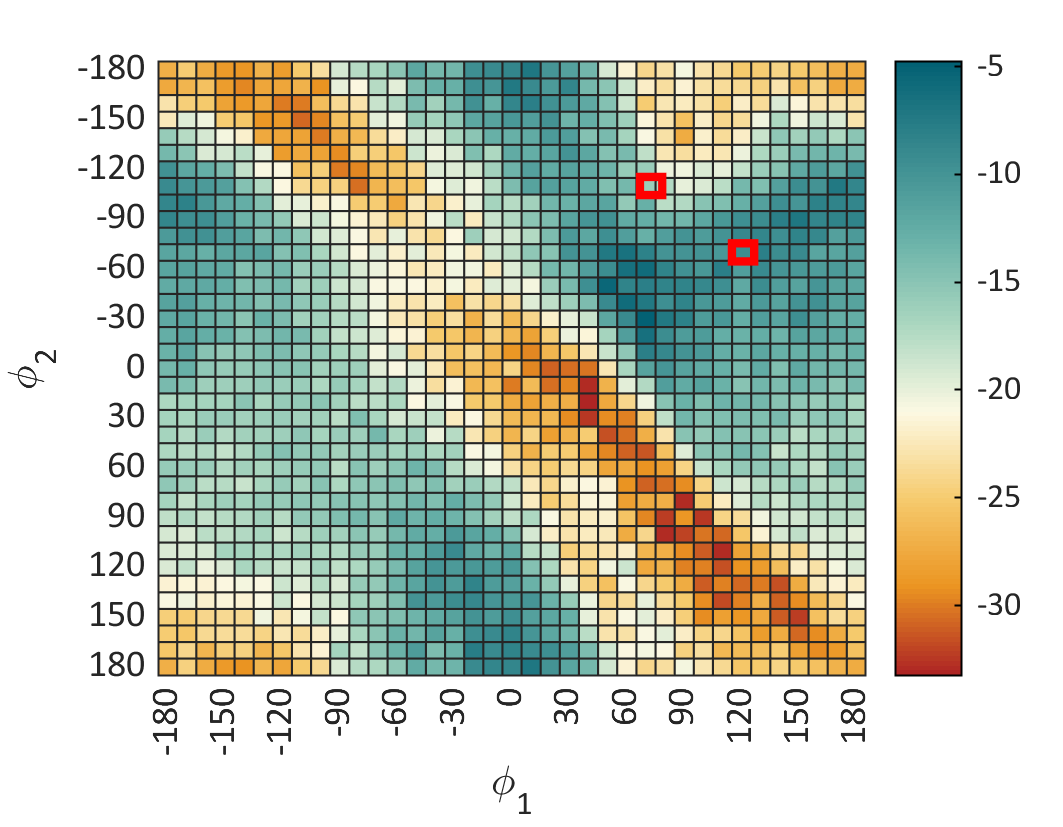}
            \label{fig:BBPs_b}
            \caption{}
        \end{subfigure}
    \caption{\label{fig:FCs} Error (in $\%$) of measured charge by large 12.5~$\Omega$ (top) and compact 50~$\Omega$ (bottom) FCs over 2D RF phase scan. Values above correspond to the charge intercepted by the front face of the FCs, estimated through particle tracking simulations with ASTRA~\cite{ASTRA}. Notice that the 50~$\Omega$ case corresponds to the sum of 6 narrow-range measurements, as indicated in Table~\ref{tab:FCparams}. $\Phi=$~(120,-70) and $\Phi=$~(70,-110) marked in red.}
\end{figure}

Error estimations of the measured e+ charge are shown in Fig.~\ref{fig:FCs} for both FC layouts and all RF phase configurations, showing a reasonably good agreement with Fig.~\ref{fig:2DScan} in the vicinity of the RF working point of interest. These error studies are based on ASTRA~\cite{ASTRA} particle tracking simulations. At $\Phi=$~(120,-70), the point for maximum captured e+ charge, the 12.5~$\Omega$ and 50~$\Omega$ FCs would read -13.6$\%$ and -9.4$\%$ with respect to the expected 1246~pC. This RF working point was studied in further detail with Geant4~\cite{G4} simulations. 60~mm-thick W blocks were considered for both FCs in order to maximize charge deposition. Notice that as for now, back scattering effects due to the high Tungsten density are disregarded as they are negligible above 10~MeV. The results obtained indicate a deposited e+ charge of 1058~pC in the 12.5~$\Omega$ FC and 1164~pC in the 50~$\Omega$ one, namely -15.0$\%$ and -6.6$\%$ with respect to 1246~pC. On the other hand, ooorer charge measurements are expected at lower energies, as particle divergence will have a greater impact in the final transverse position. For this reason, e+ charge can be underestimated by as much as -58$\%$ and -33$\%$ by the 12.5~$\Omega$ and 50~$\Omega$ FCs respectively. However, this occurs in regions with relatively small importance for the experiment. 

\subsection{Scintillators in Diagnostics Chamber}

Aside from the FCs, the diagnostics chamber will accommodate at least two additional instrumentation setups based on scintillator materials. First, the front face of the FCs will have scintillating screens that will allow cameras mounted outside of the chamber to look at the collected e+ and e- distributoins. These scintillator screens are particularly useful during beam commissioning, as counterpart of the FC signals. However, the large size of the FCs and the transverse emittance of the beam result in a very poor energy resolution, which precludes any usefulness of performing spectral measurements through these devices. The scintillator can either be a coating deposited on the face of the FCs, a screen mounted to the front of the FCs, or a free-standing screen that can be inserted in and out of the chamber. The most likely materials for the screen would be Cr-doped Alumina (Chromox), Biomax, or YAG, which have been used for scintillation in accelerators in the past \cite{JohnsonScintillators}. 

\begin{figure}[h!]
        \includegraphics[width=0.45\textwidth]{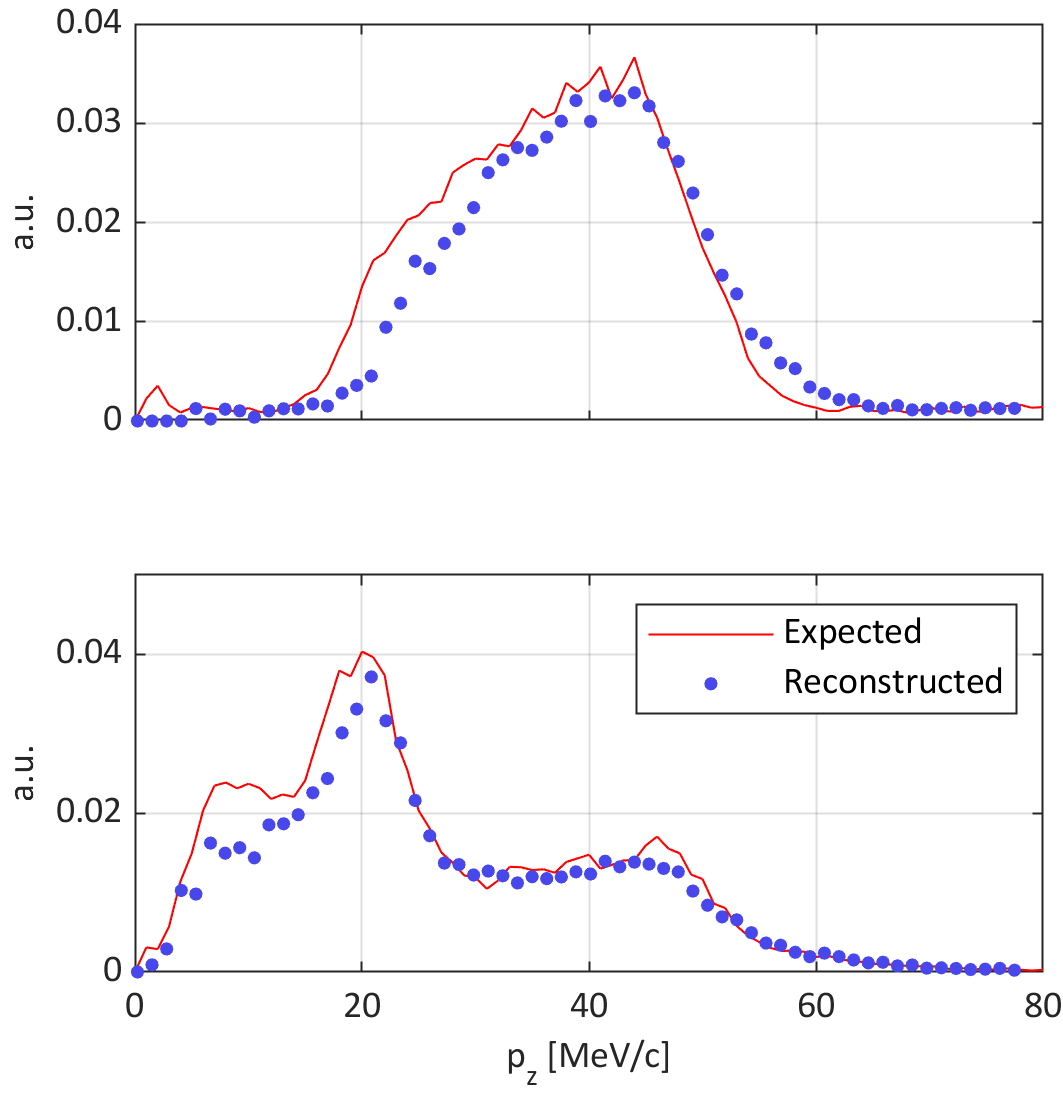}
    \caption{\label{fig:EReconstruction} Estimation of $p_z$ reconstruction for $\Phi=$~(120,-70) (top) and $\Phi=$~(70,-110) (bottom). Computed through through 61 ASTRA simulations ranging dipole field strengths from 0 to 0.3~T and scintillator fibers located at x~=~-150~mm and z~=~3520~mm.}
\end{figure}

An high-resolution spectroscopic setup consisting of at least one pair of scintillator fibers will reconstruct the longitudinal momentum ($p_z$) spectrum of the e+ and e- distributions. The fibers, vertically oriented, will be hit by a small fraction of the particles corresponding to a narrow division of the energy spectrum. Following the basic principle of beam loss monitors, scintillator fibers will emit a signal proportional to the ionisation energy released by the incident charges. This will allow to scan the dipole field strength over as many points as desired. The baseline location of the first scintillator fiber pair is considered at x~=~-150~mm and z =~3520~mm (320~mm downstream from the center of the spectrometer). Fig.~\ref{fig:EReconstruction} shows a reconstruction of the $p_z$ spectrum for both RF working points of interest ( see section~\ref{ch:BD}), computed through multiple particle tracking simulations. Despite the notoriously good agreements, the reconstruction below does not consider important factors such as scintillator response and signal acquisition. In addition, this spectral detection is supposed to be relative at the moment. Calibration of the fibers is required for absolute charge measurements and is still under investigation. 

\section{Experiment Installation}
The SwissFEL facility is an ideal host for the P$^3$ experiment since it can provide a $6$~GeV electron beam, which exactly corresponds to the nominal drive beam energy of the FCC-ee positron source (see Table~\ref{tab:FCCvsP3}).
Two beam lines (Aramis and Athos) are currently operating at SwissFEL, while the accelerator tunnel already foresees space for a future, third beam line (Porthos) leaving enough room for the installation of the P$^3$ bunker and switchyard.

\subsection{Porthos Switchyard}
The Porthos switchyard "Phase Planned" is currently being installed following the layout depicted in Fig.~\ref{fig:PorthosSwitchyardAndPCubed}, which is a simplified version of the final Porthos switchyard "Phase Future", whose design has been reported in \cite{Reiche:PorthosSwitchyard}.
The organization of the installation is particularly challenging because it can only take place during the usual machine shutdowns, respecting the nature of SwissFEL as a user facility.

A first static dipole (blue) allows to extract the beam from the Aramis to the Porthos line, meaning that the beam will be available either in the hard X-rays Aramis beam line or in the Porthos switchyard for the P$^3$ experiment.
A parallel operation of the soft X-rays Athos beam line and of the P$^3$ experiment is in principle not excluded.
Along the Porthos switchyard we further find $9$ quadrupoles, $6$ x/y correctors, $6$ beam position monitors (BPMs), $2$ beam loss monitors (BLMs) and one screen to image the beam just before entering the experimental bunker.

\subsection{High-Voltage Modulator}
\begin{figure*}
    \centering
    \includegraphics[width=\textwidth]{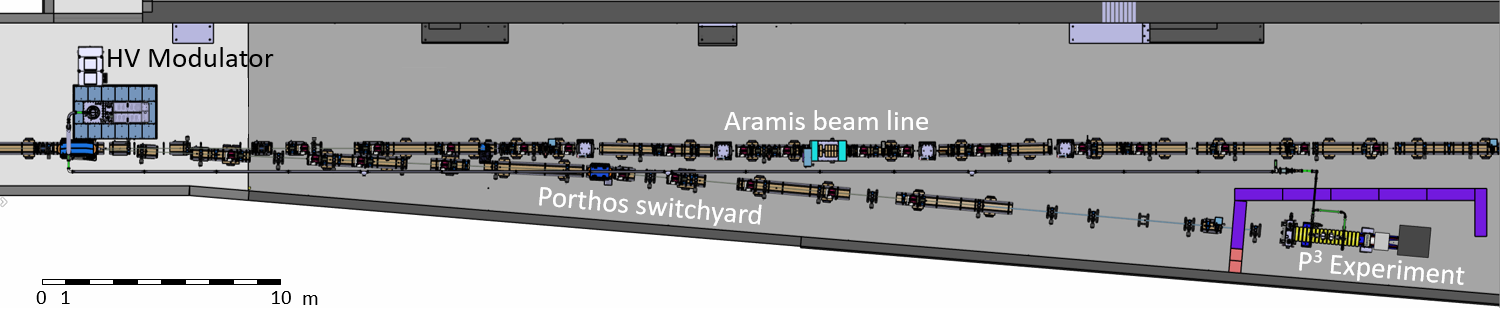}
    \caption{\label{fig:PorthosSwitchyardAndPCubed} Top view of the simplified Porthos switchyard (Phase Planned) and of the P$^3$ experiment in the SwissFEL facility. In this picture, only the existing Aramis beam line is displayed to appreciate the alternation of short girders in the first part of the new Porthos switchyard. The HV modulator (visible on the left) will be installed in the technical gallery one floor above the accelerator tunnel.}
\end{figure*}

A new HV modulator (left in Fig.~\ref{fig:PorthosSwitchyardAndPCubed}) will be installed in the last available location in the technical gallery, one floor above the accelerator tunnel at a longitudinal coordinate of $z \sim 432$~m ($z = 0$ being the emission plane of the SwissFEL photocathode gun).
With the conversion target of P$^3$ at $z \sim 483$~m, a waveguide line of about $50$~m has been projected to bring RF power in the order of $30$~MW to the S-band SW structures of the experiment.

\begin{figure*}
    \centering
    \includegraphics[width=\textwidth]{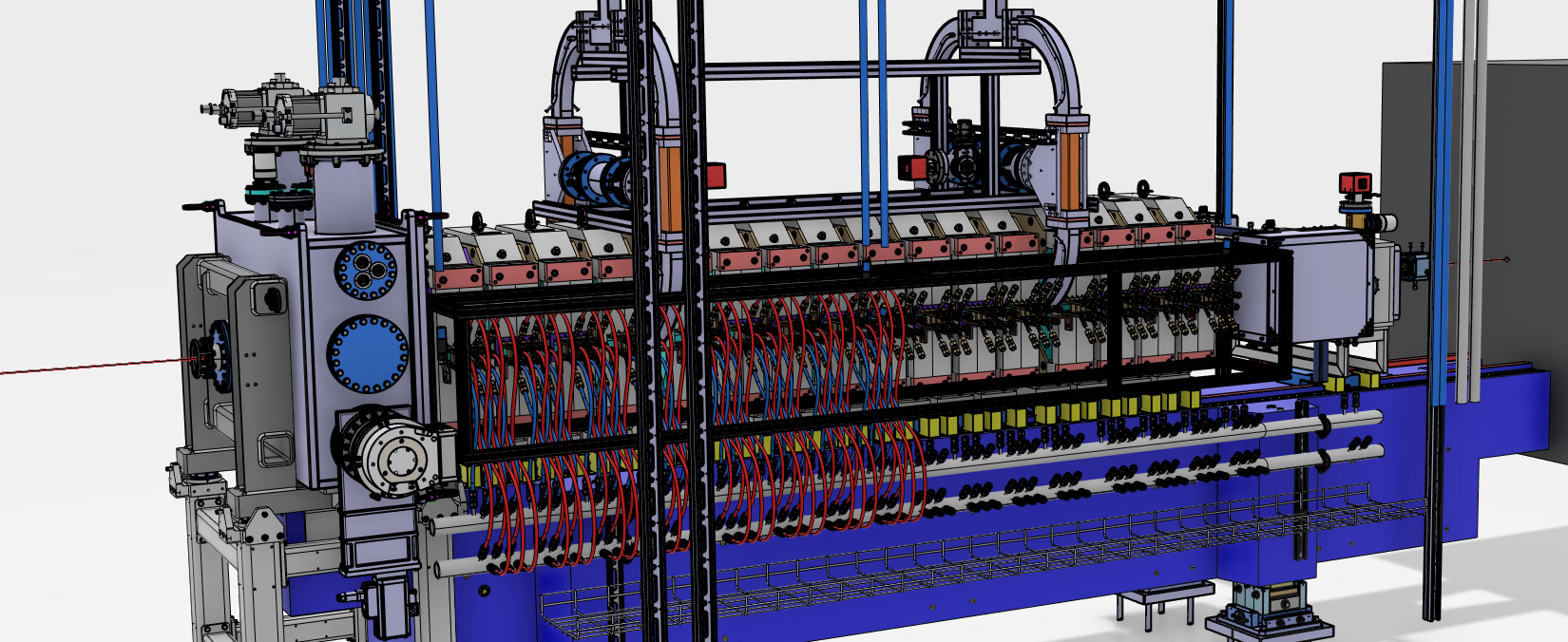}
    \caption{\label{fig:P3Render} Reconstruction of P$^3$ components mounted on a girder, recreated according to latest mechanical design assembly.}
\end{figure*}

\subsection{P$^3$ Bunker and Radiation Protection}
A bunker consisting of side walls (but no roof) of standard concrete blocks is necessary to shield from the radiation generated during the experiment, when the $200$~pC, $6$~GeV electron drive beam will impinge on the tungsten target at $1$~Hz.
The shielding has a double purpose: the respect of the legal dose limits inside and outside the facility, as well as the protection of potentially sensible machine components.

\begin{figure}
  \begin{center}
  \includegraphics[width=0.45\textwidth]{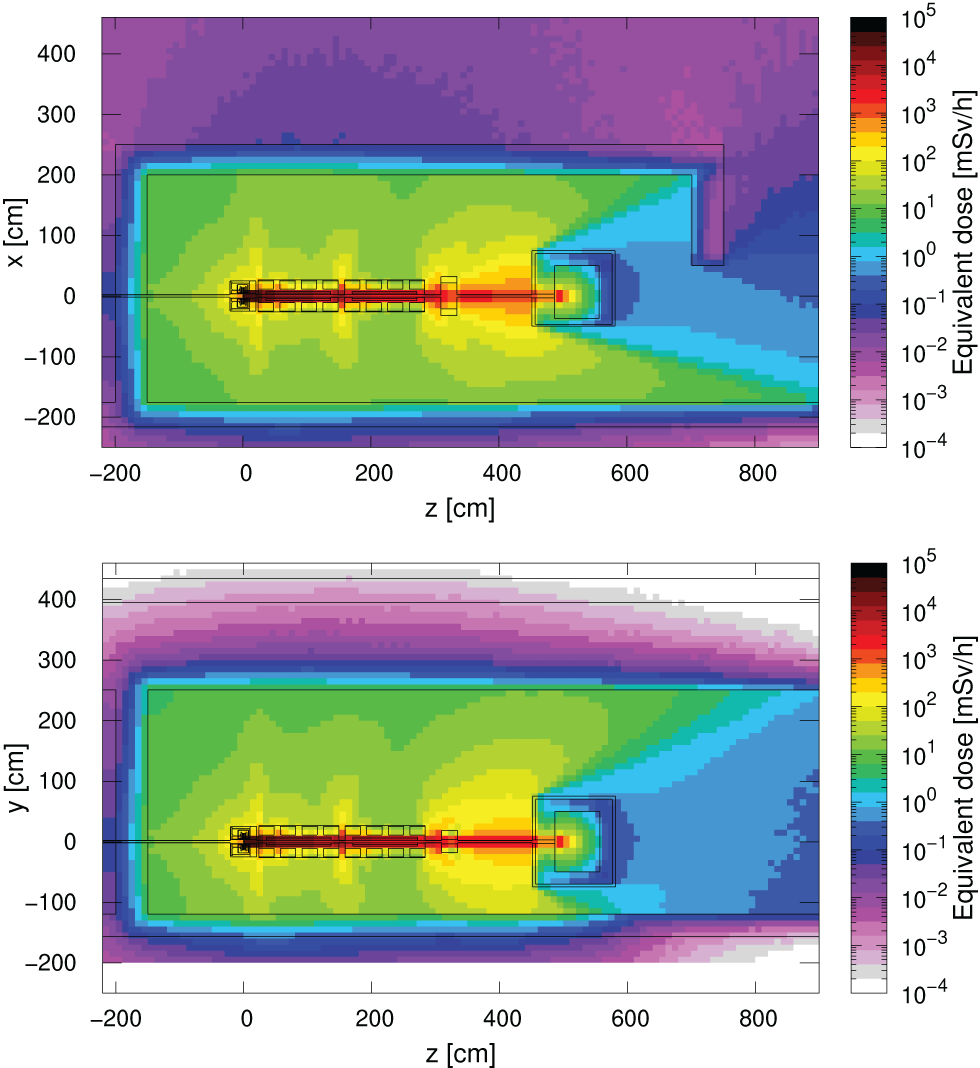}
  \end{center}
  \caption{Ambient dose equivalent in the $z-x$ plane (top) and $z-y$ plane (bottom). No filter on the particle type is applied. No Faraday cup included in the model. Final simulations with all the components in progress.}
  \label{fig:doseTarget}
\end{figure}

The radiation dose inside and outside the P$^3$ bunker is calculated through the general purpose Monte Carlo code FLUKA~\cite{Fluka1} based on a 3D model of the experiment built with Flair interface~\cite{Flair}.
Biasing techniques are used to improve the statistics behind the bunker walls.
A conservative approach was adopted to design the shielding, as the total dose is calculated by the sum of two separate simulations.
First, the interaction of the primary e- beam from SwissFEL and the W target is simulated according to the parameters in Table~\ref{tab:FCCvsP3}. In this case, the RF accelerating field and the magnetic field of the spectrometer are turned off, resulting in the distribution of Fig.~\ref{fig:doseTarget}.
Yet, a significant part of the secondary e+ and e- will be captured by the solenoid channel and are lost in the dump.
In a second simulation chain, the e-/e+ distribution is tracked with ASTRA up to the exit of the second RF structure and then imported into FLUKA, where the deflection of the spectrometer and following interaction with the diagnostic section is computed.
The dose distributions resulting from the two simulation setups are finally summed up to judge the radiation level at the relevant locations and eventually optimize the shielding.

\section{CONCLUSION} 

This paper introduced the P$^3$ experiment, a demonstrator for a highly efficient e+ source framed in the FCC-ee injector study, and based on a 6~GeV electron (e-) beam and 17.5~mm-thick (or 5$X_0$) amorphous Tungsten (W) target. The experiment design was presented at a highly advanced stage, with a particular emphasis on a novel e+ capture system consisting of a HTS solenoid, 2 large aperture RF cavities and an arrangement of 16 NC solenoids surrounding the cavities. We provided the key elements of a nearly final technical design of all the above mentioned components, including a pioneer concept and demonstration of the HTS solenoid operation at 18~T on axis. In addition, e+ yields, and transverse and longitudinal beam dynamics were examined at a fairly high level of detail through Geant4~\cite{G4} and ASTRA~\cite{ASTRA} simulations. A comprehensive beam optimization was performed based on two figures of merit: the expected captured e+ charge, and an equivalent yield provided at the FCC-ee damping ring. Accordingly, we selected two RF phase configurations as the future reference working points for optimum performance. Table~\ref{tab:Yields_p3} gathers the key figures of our studies, which indicate an increase of one order of magnitude in the e+ yield with respect to that provided by presently existing accelerators. 

\begin{table}[h]
\begin{ruledtabular}
\begin{tabular}{lcc}
		& e+ charge & Norm. to 200~pC \\
        \colrule
        Captured & 1246~pC &  6.23~$N_{e+}/N_{e-}$ \\
        Measured by FCs & 1129~pC & 5.64~$N_{e+}/N_{e-}$ \\
        At FCC-ee DR & - & 3.84~$N_{e+}/N_{e-}$ \\
        \colrule
        Captured & 1153~pC &  5.76~$N_{e+}/N_{e-}$ \\
        Measured by FCs & 971~pC & 4.86~$N_{e+}/N_{e-}$ \\
        At FCC-ee DR & - & 4.64~$N_{e+}/N_{e-}$ \\
\end{tabular}
\caption{\label{tab:Yields_p3} Summary of the key e+ yields envisaged for P$^3$}
\end{ruledtabular}
\end{table}

The proof of principle of such a yield upgrade relies on the experiment diagnostics, which would consist of an arrangement of broadband pick-ups, 2 types Faraday cups and a variety of scintillating detectors. The Faraday cups were studied in greater detail and they will arguably provide the most accurate detection of the captured e+ charge. Thus we performed a preliminary error study based on simulations, envisaging reasonably high quality measurements of the Faraday cups, specially at RF working points of interest (see Table~\ref{tab:Yields_p3}). Finally, we overviewed the ongoing installation proceedings at SwissFEL with a special emphasis on the radiation protection bunker around the experiment, the waveguide network and the new transfer line \textit{Porthos} branching from the main SwissFEL linac. 

\section{ACKNOWLEDGEMENT}
We would like to thank the great support from our colleagues in the RF section at PSI, and all the technical groups involved in the installation of the P$^3$ experiment. In addition, we would like to acknowledge the FCC collaboration, and particularly our colleagues at CERN, IJCLab and INFN involved in the FCC-ee Injector design study. This work was done under the auspices of \href{http://www.chart.ch}{CHART (Swiss Accelerator Research and Technology)}.


\clearpage
\begin{thebibliography}{99}

\bibitem{chehab:CAS}
R. Chehab, Positron Sources, \textit{in Proc. Cern Accelerator School, 5th general accelerator physics course, Jyvaskyla, Finland}, (CERN, 1994), pp. 2617-2620, \href{http://doi.org/10.5170/CERN-1994-001}{10.5170/CERN-1994-001}.

\bibitem{clendenin:positrons} 
J. E. Clendenin, High-Yield Positron Systems for Linear Colliders, in Proc. PAC'89, Chicago, USA (1989), pp. 1107–1112.

\bibitem{Chaikovska:PositronSources} I. Chaikovska, R. Chehab, V. Kubytskyi, S. Ogur, A. Ushakov, A. Variola, P. Sievers, P. Musumeci, L. Bandiera, Y. Enomoto, Mark J. Hogan and P. Martyshkin, Positron sources: from conventional to advanced  accelerator concepts-based colliders, \href{https://doi.org/10.1088/1748-0221/17/05/P05015}{JINST, \textbf{17}, P05015 (2022)}

\bibitem{SLCHandbook}
SLAC Linear Collider Design Handbook, SLAC-R-714 (1984).

\bibitem{Akai:SuperKEKB}
K. Akai, K. Furukawa and H. Koiso, SuperKEKB Collider, \href{https://doi.org/10.1016/j.nima.2018.08.017}{Nucl. Instrum. Methods
Phys. Res., Sect. A \textbf{907}, 188 (2018)}.

\bibitem{Suwada:Detection}
T. Suwada, M.A. Rehman, and F. Miyahara, First simultaneous detection of electron and positron bunches at the positron capture section of the SuperKEKB factory, \href{https://doi.org/10.1038/s41598-021-91707-0}{Sci Rep 11, 12751 (2021)}

\bibitem{Prat:SiwssFEL}
E. Prat \textit{et al.}, A compact and cost-effective hard X-ray free-electron laser driven by a high-brightness and low-energy electron beam, \href{https://doi.org/10.1038/s41566-020-00712-8}{Nat. Photonics \textbf{14}, 748–754 (2020)}. 

\bibitem{FCCweb}
Future Circular Collider study, \href{https://fcc.web.cern.ch/}{https://fcc.web.cern.ch/}.

\bibitem{FCC-ee_CDR}
M. Benedikt, F. Zimmermann \textit{et al.}, FCC-ee: the Lepton Collider. Future Circular Collider Conceptual Design Report Volume 2, \href{https://doi.org/10.1140/epjst/e2019-900045-4}{European Physical Journal Special Topics, \textbf{228(2)}, 261-623 (2019)}

\bibitem{Craievich:FCCeeInjector}
P. Craievich \textit{et al.}, The FCCee Pre-Injector Complex, \textit{in Proc. IPAC'22, Bangkok, Thailand} (JACoW, Geneva, Switzerland, 2022), pp. 2007-2010, \href{http://doi.org/10.18429/JACoW-IPAC2022-WEPOPT063}{10.18429/JACoW-IPAC2022-WEPOPT063}.

\bibitem{CHART21}
P. Craievich, M. Schaer, N. Vallis and R. Zennaro, FCC-ee Injector Study and
the P$^3$ Project at PSI, CHART Scientific Report (2021), \href{https://chart.ch/wp-content/uploads/2022/05/Chart-Scientific-Report-2021-FCCee-Injector.pdf}{https://www.chart.ch}. 

\bibitem{CHART22}
P. Craievich \textit{et al.}, FCC-ee Injector Study and
the P$^3$ Project at PSI, CHART Scientific Report (2022), \href{https://chart.ch/wp-content/uploads/2023/02/FCCee-Injector-2022.pdf}{https://www.chart.ch}. 

\bibitem{Chaikovska:FCC-ee}
I. Chaikovska \textit{et al.}, Positron source for FCC-ee, \textit{in Proc. IPAC'19, Melbourne, Australia} (JACoW, Geneva, Switzerland, 2019), pp. 424-427, \href{http://doi.org/10.18429/JACoW-IPAC2019-MOPMP003}{10.18429/JACoW-IPAC2019-MOPMP003}.



\bibitem{Zhao:AMDcomparison}
Y. Zhao \textit{et al.}, Comparison of Different Matching Device Field Profiles for the FCC-ee Positron Source, \textit{in Proc. IPAC'21, Campinas, SP, Brazil}, (JACoW, Geneva, Switzerland, 2021), pp. 2617-2620, \href{http://doi.org/10.18429/JACoW-IPAC2021-WEPAB015}{10.18429/JACoW-IPAC2021-WEPAB015}.

\bibitem{Zhao:AMDOptimization}
Y. Zhao, B. Auchmann, I. Chaikovska, R. Chehab, P. Craievich, S. Döbert,D. Duda, J. Kosse, A. Latina, P. Martyshkin, S. Ogur, R. Zennaro, Optimisation of the FCC-ee Positron Source Using a HTS Solenoid Matching Device, \textit{in Proc. IPAC'22, Bangkok, Thailand} (JACoW, Geneva, Switzerland, 2022), pp. 2003-2006. \href{http://doi.org/10.18429/JACoW-IPAC2022-WEPOPT062}{10.18429/JACoW-IPAC2022-WEPOPT062}.

\bibitem{Humann:AMDRadiation}
B. Humann \textit{et al}., Radiation Load Studies for the FCC-ee Positron Source with a Superconducting Matching Device, \textit{in Proc. IPAC'22, Bangkok, Thailand} (JACoW, Geneva, Switzerland, 2022), pp. 2879-2882. \href{http://doi.org/10.18429/JACoW-IPAC2022-THPOTK048}{doi.org/10.18429/JACoW-IPAC2022-THPOTK048}.

\bibitem{Hahn:HTSNoInsulation}
S. Hahn, D. K. Park, J. Bascunan and Y. Iwasa, HTS Pancake Coils Without Turn-to-Turn Insulation, in IEEE Trans. on Appl. Supercond., vol. 21, no. 3, pp. 1592-1595, June 2011, doi: 10.1109/TASC.2010.2093492.

\bibitem{Hahn:HTSChallenges}
D. Hahn, K. Kim, H. Lee and Y. Iwasa, Current Status of and Challenges for No-Insulation HTS Winding Technique, \href{http://doi.org/10.2221/jcsj.53.2}{Teion Kogaku: Official Journal of the Cryogenic Association of Japan, \textbf{53(1)}, 2 (2018)}.

\bibitem{Fischer:ReBCORadiation}
D.X. Fischer, R. Prokopec, J. Emhofer and M. Eisterer, The Effect of Fast Neutron Irradiation on the Superconducting Properties of REBCO Coated Conductors with and without Artificial Pinning Centers, \href{http://doi.org/10.1088/1361-6668/aaadf2}{Supercond. Sci. Technol., \textbf{31}, 044006 (2018)}

\bibitem{SumitomoCryoCoolers}
Sumitomo RDK500B, \href{https://www.shicryogenics.com/product/rdk-500b-20k-cryocooler-series/}{https://www.shicryogenics.com)}

\bibitem{G4}
Geant4: Toolkit for the Simulation of the Passage of Particles Through Matter, \href{https://geant4.web.cern.ch/}{https://geant4.web.cern.ch/}.

\bibitem{ASTRA}
ASTRA: A Space Charge Tracking Algorithm, \href{https://www.desy.de/~mpyflo/}{https://www.desy.de/~mpyflo/}.

\bibitem{Zhao:CLICOpt}
Y. Zhao, A. Latina, S. Doebert, D. Schulte and L. Ma, Optimisation of the CLIC positron source at the 1.5 TeV and 3 TeV stages, \href{https://cds.cern.ch/record/2735292}{CERN-ACC-2020-0026, CLIC-Note-1165}.

\bibitem{floettmann:emit}
K. Floettmann, Some basic features of the beam emittance, \href{https://doi.org/10.1103/PhysRevSTAB.6.034202}{Phys. Rev. Accel. Beams \textbf{6}, 034202 (2003)}.

\bibitem{Helm:AMD}
R.H. Helm, Adiabatic Approximation for Dynamics of a Particle in the Field of a Tapered Solenoid, SLAC Report No. 4 (1962).

\bibitem{chehab:CAS2}
R. Chehab, Positron Sources, \textit{in Proc. Cern Accelerator School, 3rd general accelerator physics course, Salamanca, Spain}, (CERN, 1988), pp. 105-132, \href{http://doi.org/10.5170/CERN-1989-005}{10.5170/CERN-1989-005}.


\bibitem{Vallis:Linac22}
N. Vallis \textit{et al.}, The PSI Positron Production Project, \textit{in Proc. LINAC'22, Liverpool, UK} (JACoW, Geneva, Switzerland, 2023), pp. 577-580. \href{http://doi.org/10.18429/JACoW-LINAC2022-TUPORI16}{doi:10.18429/JACoW-LINAC2022-TUPORI16}

\bibitem{Aune-Miller:AMD}
B. Aune, R.H. Miller, New Method for Positron Production at SLAC, SLAC-PUB-2393, (1979).


\bibitem{CST}
CST Studio Suite Electromagnetic Simulation Solvers, \href{https://www.3ds.com/products-services/simulia/products/cst-studio-suite/solvers/}{https://www.3ds.com}

\bibitem{Allectra27}
27GHz SMA Feedthrough for UHV Applications 242-SMAD27G-C16, Allectra GmbH (2019).

\bibitem{Allectra65}
Microwave 1.85mm (SMA) feedthrough 242-SMAD65G-C16, Allectra GmbH (2022).

\bibitem{PasternackCables}
1.85mm Male to 3.5mm Male Cable Assemblies, Pasternack Enterprises, \href{https://www.fluka.org}{https://www.pasternack.com/1.85mm-male-to-3.5mm-male-cable-assemblies-category.aspx}

\bibitem{SuwadaBPM1}
T. Suwada, M.A. Rehman, and  F. Miyahara, First simultaneous detection of electron and positron bunches at the positron capture section of the SuperKEKB factory, \href{https://doi.org/10.1038/s41598-021-91707-0}{Sci. Rep. \textbf{11}, 12751 (2021)}

\bibitem{SuwadaBPM2}
T. Suwada, Direct observation of positron capture process at the positron source of the superKEKB B-factory, \href{https://doi.org/10.1038/s41598-022-22030-5}{Sci. Rep. \textbf{12}, 18554 (2022).}

\bibitem{SwissFELBAM}
A.~Angelovski, A.~Kuhl, M.~Hansli, A.~Penirschke, S.~M.~Schnepp, M.~Bousonville, H.~Schlarb, M.~K.~Bock, T.~Weiland and R.~Jakoby, High Bandwidth Pickup Design for Bunch Arrival-time Monitors for Free-Electron Laser, \href{https://doi.org/10.1103/PhysRevSTAB.15.112803}{Phys. Rev. Accel. Beams \textbf{15}, 112803 (2012)}.

\bibitem{HFSS}
Ansys HFSS, Best-In-Class 3D High Frequency Structure Simulation Software, \href{https://www.ansys.com/products/electronics/ansys-hfss}{https://www.ansys.com}

\bibitem{JohnsonScintillators}
C.D. Johnson, The development and use of alumina ceramic fluorescent screens, CERN Report No. CERN-PS-90-42-AR (1990).



\bibitem{Reiche:PorthosSwitchyard}
S.\ Reiche \textit{et al.}, Design Considerations for the Extraction Line of the Proposed Third Beamline Porthos at SwissFEL, \textit{in Proc. FEL'22, Trieste, Italy}, WEP09, \href{https://indico.jacow.org/event/44/contributions/532}{indico.jacow.org/event/44/contributions/532}.

\bibitem{Fluka1} The Official FLUKA Site, \href{https://www.fluka.org}.

\bibitem{Flair} V. Vlachoudis, FLAIR: A Powerful But User Friendly Graphical Interface For FLUKA, in \textit{Proc. Int. Conf. on Mathematics, Computational Methods \& Reactor Physics, Saratoga Springs, New York, 2009}, pp.790-800.





  
\end{thebibliography}
\end{document}